\documentclass[twocolumn]{aastex63}
\usepackage{bm,amsmath} 

\graphicspath{{./}{figures/}}

\newcommand{\prospector}{\texttt{Prospector}}

\newcommand{\mcomp}{M_{\mathrm{c}}}

\begin{document}

\title{A New Census of the $\bm{0.2 < z < 3.0}$ Universe, Part I: The Stellar Mass Function}

\correspondingauthor{Joel Leja}
\email{joel.leja@cfa.harvard.edu}

\author[0000-0001-6755-1315]{Joel Leja}
\affil{Center for Astrophysics\:$|$\:Harvard \& Smithsonian, 60 Garden St. Cambridge, MA 02138, USA}

\author[0000-0003-2573-9832]{Joshua S. Speagle}
\affil{Center for Astrophysics\:$|$\:Harvard \& Smithsonian, 60 Garden St. Cambridge, MA 02138, USA}

\author[0000-0002-9280-7594]{Benjamin D. Johnson}
\affil{Center for Astrophysics\:$|$\:Harvard \& Smithsonian, 60 Garden St. Cambridge, MA 02138, USA}

\author[0000-0002-1590-8551]{Charlie Conroy}
\affil{Center for Astrophysics\:$|$\:Harvard \& Smithsonian, 60 Garden St. Cambridge, MA 02138, USA}

\author[0000-0002-8282-9888]{Pieter van Dokkum}
\affil{Department of Astronomy, Yale University, New Haven, CT 06511, USA}

\author[0000-0002-8871-3026]{Marijn Franx}
\affil{Leiden Observatory, Leiden University, NL-2300 RA Leiden, Netherlands}

\submitjournal{ApJ}
\begin{abstract}
There has been a long-standing factor-of-two tension between the observed star formation rate density and the observed stellar mass buildup after $z\sim2$. Recently we have proposed that sophisticated panchromatic SED models can resolve this tension, as these methods infer systematically higher masses and lower star formation rates than standard approaches. In a series of papers we now extend this analysis and present a complete, self-consistent census of galaxy formation over $0.2 < z < 3$ inferred with the \prospector{} galaxy SED-fitting code. In this work, Paper I, we present the evolution of the galaxy stellar mass function using new mass measurements of $\sim$10$^5$ galaxies in the 3D-HST and COSMOS-2015 surveys. We employ a new methodology to infer the mass function from the observed stellar masses: instead of fitting independent mass functions in a series of fixed redshift intervals, we construct a continuity model that directly fits for the redshift evolution of the mass function. This approach ensures a smooth picture of galaxy assembly and makes use of the full, non-Gaussian uncertainty contours in our stellar mass inferences. The resulting mass function has higher number densities at a fixed stellar mass than almost any other measurement in the literature, largely owing to the older stellar ages inferred by \prospector{}. The stellar mass density is $\sim$50\% higher than previous measurements, with the offset peaking at $z\sim1$. The next two papers in this series will present the new measurements of star-forming main sequence and the cosmic star formation rate density, respectively.
\end{abstract}

\section{Introduction}

Galaxies acquire their stars through a combination of in-situ star formation and merging with other galaxies. This growth is difficult to simulate from first principles as it requires modeling a wide range of processes on physical scales from stellar to cosmological (e.g., {Somerville} \& {Dav{\'e}} 2015). Observations of the stellar mass function are thus a critical constraint for hydrodynamical, empirical, and analytical models of galaxy formation (e.g., {Lilly} {et~al.} 2013; {Genel} {et~al.} 2014; {Furlong} {et~al.} 2015; {Somerville} \& {Dav{\'e}} 2015; {Pillepich} {et~al.} 2018; {Grylls} {et~al.} 2019; {Behroozi} {et~al.} 2019; {Dav{\'e}} {et~al.} 2019; {Grylls} {et~al.} 2020). Accordingly, accurate measurements of the stellar mass function have been a subject of intense observational interest ({Marchesini} {et~al.} 2009; {Muzzin} {et~al.} 2013; {Ilbert} {et~al.} 2013; {Moustakas} {et~al.} 2013; {Tomczak} {et~al.} 2014; {Grazian} {et~al.} 2015; {Song} {et~al.} 2016; {Davidzon} {et~al.} 2017; {Wright}, {Driver}, \& {Robotham} 2018).

Stellar masses are inferred from observations by constructing models for the combined emission of the physical components of galaxies, including stars, gas, dust, and supermassive black holes, and fitting them to the observed galaxy photometry (see, e.g., the review by {Conroy} 2013). Typically, these spectral energy distribution (SED) models consist of a combination of stellar templates, prescriptions for dust physics, and a minimization routine (e.g. \texttt{FAST}, {Kriek} {et~al.} 2009, \texttt{Le Phare}, {Arnouts} {et~al.} 1999; {Ilbert} {et~al.} 2006, and \texttt{MAGPHYS}, {da Cunha}, {Charlot}, \&  {Elbaz} 2008). Recently a new generation of these codes have emerged which allow the creation of more complex models generated on-the-fly, including \texttt{BayeSED} ({Han} \& {Han} 2014), \texttt{BEAGLE} ({Chevallard} \& {Charlot} 2016), \prospector{} ({Leja} {et~al.} 2017; Johnson \& Leja 2017), and \texttt{BAGPIPES} ({Carnall} {et~al.} 2019). These codes permit much more model flexibility, allowing users to relax many of the strong assumptions which typically go into these fits.

Using \prospector{}, {Leja} {et~al.} (2019b) fit the rest-frame UV-IR photometry of a large sample of galaxies at $0.5 < z < 2.5$ from the 3D-HST photometric catalogs ({Skelton} {et~al.} 2014; {Momcheva} {et~al.} 2016). Relative to previous methodologies, this study inferred stellar masses which are systematically larger by $0.1-0.3$ dex and star formation rates (SFRs) which are systematically lower by $\sim0.1-1$ dex or more. These offsets are a result of the inclusion of a wider range of physics. The dominant causes of these offsets are the substantially older stellar ages inferred with nonparametric star formation histories ({Carnall} {et~al.} 2019; {Leja} {et~al.} 2019a), and the fact that we self-consistently account for the light from old stars in the SFR inferences (see {Leja} {et~al.} (2019b)). Importantly, these offsets imply a $\sim$0.2 dex decrease in the cosmic star formation rate density and a $\sim 0.2$ dex increase in the derivative of the cosmic stellar mass density. If correct, this finding removes a long-standing factor of two disagreement between these quantities ({Madau} \& {Dickinson} 2014; {Leja} {et~al.} 2015; {Tomczak} {et~al.} 2016; {Katsianis}, {Tescari}, \&  {Wyithe} 2016; {Davidzon} {et~al.} 2018). 

However, {Leja} {et~al.} (2019b) estimated the cosmic star formation rate and stellar mass densities by applying offsets to existing measurements of the stellar mass function and star-forming sequence. This approach neglects a number of second-order effects in the determination of these integrated quantities, such altered shapes for these functions and object-by-object scatter. A full cosmic census coupled with the appropriate volume and completeness corrections is necessary to complete the picture implied by {Leja} {et~al.} (2019b). 

This paper is the first of a series of three papers which follow up {Leja} {et~al.} (2019b) by re-measuring the stellar mass function, the star-forming sequence, and inferring the new star formation rate density and rate of galaxy assembly implied by the \prospector{} results. In this work, Paper I, we use stellar masses inferred with \prospector{} to constrain the stellar mass function between $0.2 < z < 3$. The fits have been performed to publicly available photometry and redshifts from the 3D-HST ({Skelton} {et~al.} 2014) and COSMOS-2015 ({Laigle} {et~al.} 2016) catalogs. 

We introduce a new methodology for fitting the galaxy stellar mass function. This new methodology is an extension of the maximum likelihood method introduced by {Sandage}, {Tammann}, \& {Yahil} (1979). Previously, the standard approach fit separate stellar mass functions to galaxies in discrete redshift bins. The growth of the stellar mass function is then inferred by comparing the mass functions inferred at different redshifts. The main drawback to this approach is that the resulting mass functions are not guaranteed to evolve smoothly or even monotonically with redshift (e.g., {Drory} {et~al.} 2009; {Leja} {et~al.} 2015; {Tomczak} {et~al.} 2016). This uneven evolution can be caused by effects such as fluctuations in the density field due to large-scale cosmic structures or by the well-known degeneracies in the fitting functions typically used for the stellar mass function.

Instead, our new methodology fits a smooth model to the redshift evolution of the stellar mass function which is constrained simultaneously by every galaxy in the survey. The underlying assumption is that the mass functions in adjacent volumes smoothly evolve into one another. This assumption makes this approach more robust to both fluctuations in the density field and degeneracies in the fitting functions.

The photometric data and redshifts are described in Section \ref{sec:data} and the SED modeling is described in Section \ref{sec:sedmodel}. The mass function model is described in Section \ref{sec:mf_model}. The results are presented in Section \ref{sec:results}. Section \ref{sec:discussion} discusses the broader context of these results and the conclusion is presented in Section \ref{sec:conclusion}. We use a {Chabrier} (2003) initial mass function and adopt a WMAP9 cosmology ({Hinshaw} {et~al.} 2013) with $H_0=69.7$ km/s/Mpc, $\Omega_b = 0.0464$, and $\Omega_c=0.235$. Parameters are reported as the median of the posterior probability distribution functions and uncertainties are half of the (84$^{\mathrm{th}}$-16$^{\mathrm{th}}$) percentile range, unless indicated otherwise.

\section{Data}
\label{sec:data}

Here we describe the photometry, redshifts, and areal coverage from the surveys used in this work. These data are all taken from publicly available catalogs.

\subsection{3D-HST}
\label{sec:3dhst}

The 3D-HST photometric catalogs cover five well-studied extragalactic fields with a total area of $\sim900$ arcmin$^2$ ({Skelton} {et~al.} 2014). The provided photometry ranges from 17 to 44 bands and spans 0.3-8$\mu$m in the rest-frame. It is supplemented with Spitzer/MIPS photometry from {Whitaker} {et~al.} (2014). Crucially, the fields include deep HST imaging from the CANDELS program ({Grogin} {et~al.} 2011; {Koekemoer} {et~al.} 2011). The survey also provides measured redshifts; for the objects fit in this work, approximately 30\% are measured spectroscopic or grism redshifts ({Momcheva} {et~al.} 2016) while the remaining $\sim$70\% are photometric redshifts from EAZY ({Brammer}, {van Dokkum}, \&  {Coppi} 2008).

We adopt \prospector{} fits to this catalog from {Leja} {et~al.} (2019b), which include 58,461 galaxies selected above the stellar mass completeness limit between $0.5 < z < 2.5$. This is done in order to limit the computational demands of running the \prospector{} model. This sample is supplemented with 4,966 objects fit with the same model between $2.5 < z < 3.0$ to extend the analysis to higher redshifts, for a total of 63,427 objects. The photometric zero-points and uncertainties are adjusted from the default 3D-HST catalog as described in {Leja} {et~al.} (2019b).

Accurate measurements of the mass function also require an accurate estimate of the mass-completeness limit $\mcomp{}(z)$, defined as the lowest stellar mass at which the galaxy sample is 100\% complete. In this work $\mcomp{}$ is set by computational constraints rather than magnitude limits, in the sense that there were only computational resources to fit a fraction of the full photometric catalogs with \prospector{}. Here we choose to fit objects down to the mass-complete limit of the 3D-HST survey as determined by {Tal} {et~al.} (2014). This selection is determined using stellar masses from the \texttt{FAST} SED-fitting code ({Kriek} {et~al.} 2009).

\begin{figure*}[t!h!]
\begin{center}
\includegraphics[width=0.8\linewidth]{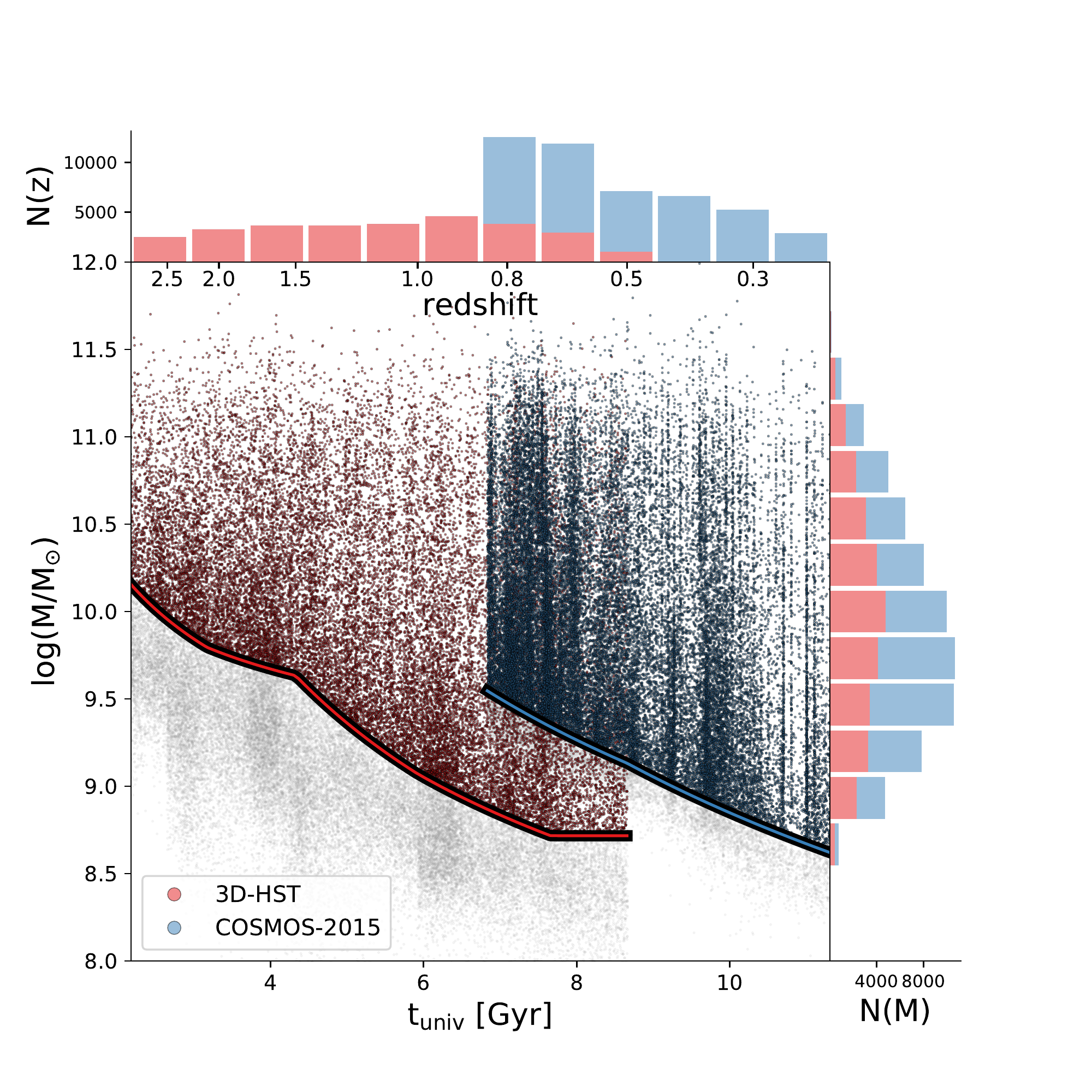}
\caption{The distribution in mass and redshift for objects from the 3D-HST and COSMOS-2015 surveys. The thick lines indicate mass-complete limits, largely set by sub-sampling of the full catalog. Grey objects are below the mass-complete limit. The vertical striping comes from large-scale cosmic structure.}
\label{fig:surveys}
\end{center}
\end{figure*}

This is not necessarily straightforward to interpret, as \texttt{FAST} stellar masses have both substantial scatter with, and are substantially offset from, the \prospector{} stellar masses ({Leja} {et~al.} 2019b).  Accordingly, to determine a stellar mass completeness limit for the \prospector{} analysis, we first correct the measured \texttt{FAST} mass completeness limits for the systematic offset between \prospector{} and \texttt{FAST}. We then add twice the measured Gaussian scatter between the two mass measurements. This calculation is performed iteratively, taking care to ensure that stellar mass incompleteness affects neither the bias nor the scatter measurements. The resulting galaxy sample and stellar mass limits are shown in Figure \ref{fig:surveys}, and the stellar mass limits are tabulated in Table \ref{table:mlim}.

\begin{deluxetable}{cc}
\tablecolumns{2}
\tablecaption{Mass completeness limits for the \prospector{} fits to the 3D-HST and COSMOS-2015 surveys \label{table:mlim}}
\tablehead{
\colhead{redshift} & \colhead{log$_{10}$(M$_{*,\mathrm{complete}}$/M$_{\odot}$)}
}
\startdata
\cutinhead{3D-HST survey}
0.65 & 8.72 \\
1.0 & 9.07 \\
1.5 & 9.63 \\
2.1 & 9.79 \\
3.0 & 10.15 \\
\cutinhead{\centering COSMOS-2015 survey}
0.175 & 8.58 \\
0.5 & 9.13 \\
0.8 & 9.55
\enddata
\end{deluxetable}

\subsection{COSMOS-2015}
We also fit objects in the COSMOS-2015 photometric catalog ({Laigle} {et~al.} 2016). This catalog contains roughly half a million objects from the 2 deg$^2$ COSMOS field ({Laigle} {et~al.} 2016), with photometry covering the rest-frame UV to the mid-infrared (including the far-infared for $<1\%$ of objects). The survey also provides measured redshifts; these redshifts are from a mixture of spectroscopic and photometric data. Importantly, COSMOS-2015 provides the volume necessary to measure the evolution of the mass function down to $z=0.2$. It also overlaps with the redshift range of the 3D-HST sample, providing a useful consistency check between the two surveys.

We select objects from the COSMOS-2015 catalog in the overlap between the COSMOS and UltraVISTA surveys ({McCracken} {et~al.} 2012) which have reliable optical photometry (i.e., FLAG\_PETER=0 in the catalog notation). The UltraVISTA survey provides the deep near-infrared photometry crucial for accurate stellar mass measurements. This overlap corresponds to a reduced area of 1.38 deg$^2$ ({Laigle} {et~al.} 2016). We further filter for objects with $0.2<z<0.8$ and M$_{\mathrm{Laigle}} > $ M$_{\mathrm{complete}}$(z), for a total of 48,443 targets. The upper redshift limit ensures overlap with the 3D-HST redshift while the lower limit avoids the saturation limit for bright, nearby galaxies ({Davidzon} {et~al.} 2017). 

The mass completeness is estimated with the same methodology described in Section \ref{sec:3dhst}, with masses and mass completeness limits taken  from the {Laigle} {et~al.} (2016) catalog. The galaxy sample and stellar mass completeness is shown in Figure \ref{fig:surveys} and the mass completeness is tabulated in Table \ref{table:mlim}.

\section{SED Modeling}
\label{sec:sedmodel}
We use the galaxy SED-fitting code \prospector{} to fit the photometry. \prospector{} infers galaxy properties using stellar populations generated by the Flexible Stellar Population Synthesis (FSPS) code ({Conroy}, {Gunn}, \& {White} 2009). The MIST stellar evolutionary tracks and isochrones ({Choi} {et~al.} 2016; {Dotter} 2016) from the MESA open-source stellar evolution package ({Paxton} {et~al.} 2011, 2013, 2015, 2018) are taken as stellar models.

We use the \prospector{}-$\alpha$ model {Leja} {et~al.} (2019b), a modified version of the model from {Leja} {et~al.} (2017). The model has 14 parameters, including a seven-component nonparametric star formation history, a two-component dust attenuation model with a flexible dust attenuation curve, free gas-phase and stellar metallicity, and mid-infrared emission from a dust-enshrouded AGN ({Leja} {et~al.} 2018). It includes dust heating from stellar sources via energy balance, emitted into a dust SED of fixed shape ({Draine} \& {Li} 2007). \prospector{} includes a self-consistent nebular emission model whereby the gas is ionized by the same stars synthesized in the SED ({Byler} {et~al.} 2017).

\begin{figure*}[t!h!]
\begin{center}
\includegraphics[width=0.95\linewidth]{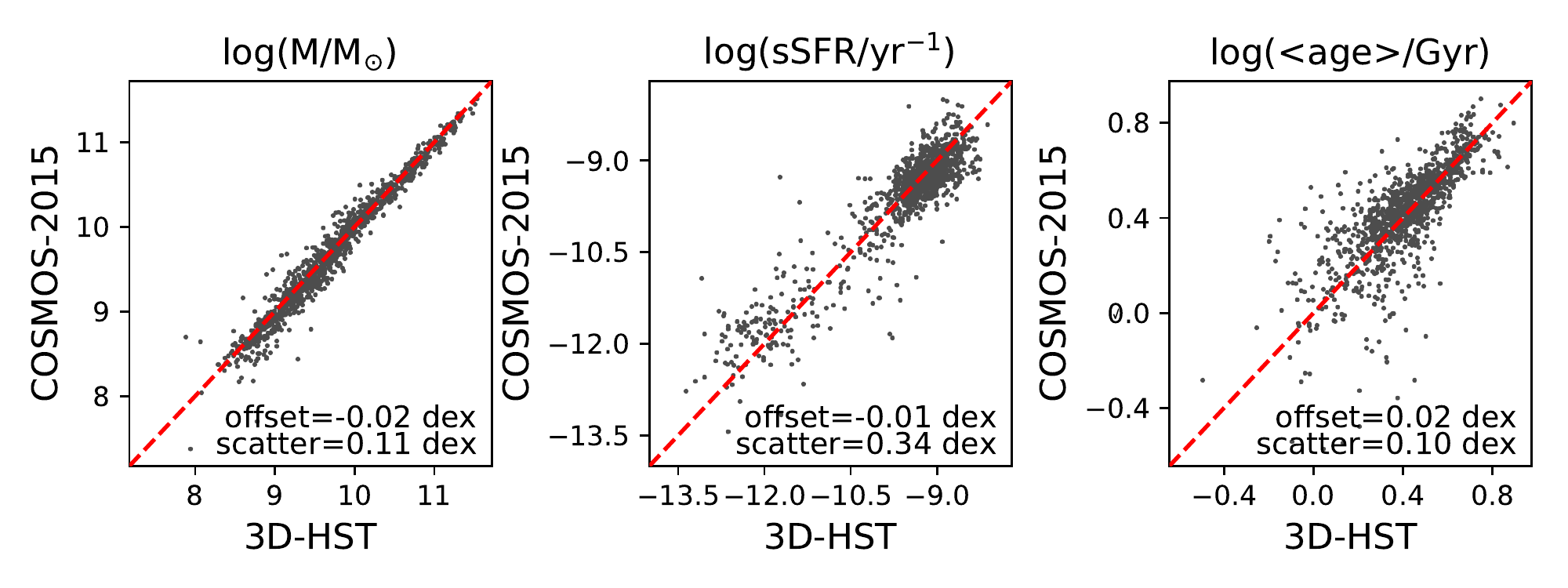}
\caption{Comparing SED-derived quantities for overlapping objects between the COSMOS-2015 and 3D-HST samples. From left to right, the properties are stellar mass, specific star formation rate, and mass-weighted age. This demonstrates that any existing photometric differences between the two catalogs do not strongly affect the SED-derived parameters.}
\label{fig:run_comparison}
\end{center}
\end{figure*}

For consistency, the same model is used to fit both COSMOS-2015 and 3D-HST. There are $\sim$1100 galaxies which overlap between the COSMOS-2015 and 3D-HST samples, matching objects are identified with a 0.2$''$ positional match and $\delta z < 0.01$. This overlap is used to explore the robustness of the SED-derived parameters to photometry measured by different teams. Figure \ref{fig:run_comparison} compares the derived parameters for the same objects. The offsets are $\lesssim 0.02$ dex, suggesting that the continuity model can be fit to both surveys without introducing substantial systematic offsets.

\section{A continuity model for the stellar mass function}
\label{sec:mf_model}
Here, we motivate and describe our continuity modeling approach for measuring the stellar mass function.

\subsection{Overview}
\label{sec:overview}
There are two standard approaches in the literature to fitting the stellar mass function.  The first is the 1/V$_{max}$ method, originally defined in {Schmidt} (1968) and later refined in {Avni} \& {Bahcall} (1980). This approach calculates the number density of objects in bins of stellar mass with
\begin{equation}
    n = N / V_{max}
\end{equation}
where $N$ is the observed number of objects and V$_{max}$ is the maximum volume out to which these objects could be detected. This calculation makes no {\it a priori} assumptions about the shape of the mass function. This approach has the advantage of flexibility, at the cost of being more sensitive to density fluctuations ({Marchesini} {et~al.} 2007) and providing no functional form for extrapolation.

The second approach is the maximum likelihood method ({Sandage} {et~al.} 1979). This is a parametric maximum likelihood estimator which assumes some functional form for the stellar mass function, typically a Schechter function ({Schechter} 1976). Deep measurements of the mass function often find that using two Schechter functions provides a better fit to the data, particularly at $z < 2$ (e.g., {Baldry}, {Glazebrook}, \&  {Driver} 2008; {Moustakas} {et~al.} 2013; {Ilbert} {et~al.} 2013; {Muzzin} {et~al.} 2013; {Tomczak} {et~al.} 2014; {Davidzon} {et~al.} 2017; {Wright} {et~al.} 2018). Fundamentally, this method assumes that the mass function $\Phi(M)$ has a universal form separable into a function of mass multiplied by density, i.e. N(M,$\bm{x}$) = $\Phi(M) \rho(\bm{x})$. This makes the fitting results robust to density inhomogeneities ({Efstathiou}, {Ellis}, \&  {Peterson} 1988). Additionally, it requires no binning in stellar mass, and can easily be extrapolated beyond the observed limits. The disadvantage relative to the 1/V$_{max}$ method is the assumption of a parametric form, effectively imposing a shape prior which can bias the resulting mass functions.

To infer the stellar mass function in a survey between some redshifts z$_{min}$ to z$_{max}$, the survey is typically split into multiple discrete redshift bins and independent mass functions are fit in each redshift interval using one of the above techniques. The evolution of the stellar mass function is then inferred by calculating the change in the observed mass function between redshifts. This approach is standard in the literature ({Baldry} {et~al.} 2008; {Marchesini} {et~al.} 2009; {Moustakas} {et~al.} 2013; {Ilbert} {et~al.} 2013; {Muzzin} {et~al.} 2013; {Tomczak} {et~al.} 2014; {Davidzon} {et~al.} 2017).

The primary drawback to this methodology is the assumed independence of the mass functions in different redshift intervals. Because they are assumed to have no relation to one another, the independently-measured mass functions are not guaranteed to evolve smoothly or even monotonically with redshift. One cause of this non-monotonic evolution is density inhomogeneities: a positive fluctuation followed by a negative fluctuation can result in negative evolution with cosmic time. Other causes are the significant degeneracies in the double Schechter function between M$_*$ and the low-mass slopes $\alpha$. This can produce significant inconsistencies between even mild extrapolations of the stellar mass function below the mass completeness limit ({Drory} {et~al.} 2009; {Leja} {et~al.} 2015). This lack of consistency can cause challenges when comparing with models. Another drawback is that this approach neglects redshift evolution of the mass function {\it within} a bin: this can be especially important when computing second-order statistics such as scatter, or when using relatively wide redshift bins (e.g. {Speagle} {et~al.} 2014)

Here we take a different approach, constructing a continuity model for the redshift evolution of the stellar mass function. This model overcomes the described limitations by fitting all objects at once, using no binning in either redshift or mass. This design assumes that the mass functions at two redshifts $z_1$ and $z_2$ are linked, insofar as one smoothly evolves into the other. This is similar to earlier works which assume some evolution in the galaxy luminosity function ({Lin} {et~al.} 1999; {Blanton} {et~al.} 2003; {Andreon} 2004, 2006). This smoothness assumption also occurs in previous works which fit smooth functions to the evolution of the mass function parameters {\it after} they have been independently derived (e.g., {Drory} {et~al.} 2009; {Leja} {et~al.} 2015; {Williams} {et~al.} 2018; {Wright} {et~al.} 2018), but here we incorporate this assumption explicitly into the fit.

Furthermore, this continuity model properly accounts for uncertainties in the derived stellar masses of individual galaxies, using the full stellar mass posteriors from the SED-fitting routine. This does not require assuming Gaussian uncertainties. Forward-modeling the mass function using the full mass uncertainty budget also naturally avoids the Eddington bias ({Eddington} 1913), assuming that the derived mass uncertainties are reliable.

\subsection{Deriving the continuity model}
\label{sec:population}
Below we construct the continuity model with parameters $\bm{\rho}$ conditioned on our data $\bm{D}$. This approach is very similar to a Bayesian hierarchical model; the primary piece missing is that the mass posteriors of individual galaxies are not modified using the derived mass function. 

In brief, the input to the modeling is the set of all galaxies above the mass-complete limit. Each galaxy has a mass and a redshift: the uncertainty in the mass is given by the full probability distribution function, while the uncertainty in the redshift is ignored\footnote{We note that it is straightforward to generalize this methodology to include redshift uncertainties.} (the effect of redshift uncertainties on the stellar mass uncertainties is discussed in Section \ref{sec:mass_errors}). All galaxies are fit at the same time. The model has eleven parameters, which, in combination, completely describe the redshift evolution of the stellar mass function. It includes one additional parameter to describe the sampling variance induced by large-scale cosmic structure. The redshift evolution is parameterized such that the evolution is smooth -- though not necessarily monotonic -- at all masses. 

The formalism follows below. A test of the formalism using mock data is shown in Appendix \ref{sec:mock}. Readers primarily interested in the modeling choices may skip the equations in both Sections \ref{sec:population} and \ref{sec:cv} without loss of clarity. 

By Bayes' Theorem:
\begin{equation}
    P(\bm{\rho} | D) = \frac{P(\bm{D} | \bm{\rho}) P(\bm{\rho})}{P(\bm{D})}
\end{equation}
The bold-face denotes vector quantities. $P(\bm{D})$ is a normalizing constant which we ignore here, and P$(\bm{\rho})$ are the priors.

The most important term is the likelihood, $P(\bm{D}|\bm{\rho})$. Here we model the redshift evolution of the galaxy stellar mass function as a Poisson point process with some occurrence rate $\lambda$. While typically $\lambda$ is taken to be fixed, here the Poisson process operates over a redshift range in which the number density of galaxies undergoes significant evolution. We therefore consider an inhomogeneous Poisson process where the rate $\lambda$ is a function of both the logarithmic mass $\mathcal{M} \equiv \log_{10}(M)$ and redshift $z$.

Ignoring constants, the probability density function for an inhomogeneous Poisson point process in $\mathcal{M}$ and $z$ with $N$ observations $\{(\mathcal{M}_1,z_1), \dots, (\mathcal{M}_N, z_N)\}$ is
\begin{equation} \label{eqn_poisson}
    P(\{(\mathcal{M}_1,z_1), \dots, (\mathcal{M}_N, z_N)\}) = e^{-N_\lambda} \prod_{i=1}^N \lambda(\mathcal{M}_i,z_i)
\end{equation}
where $N_\lambda$ is defined as
\begin{equation} \label{eqn_nlambda}
    N_\lambda \equiv \int_{z_{l}}^{z_{h}}\int_{\mcomp}^{\mathcal{M}_h} \lambda(\mathcal{M},z) d\mathcal{M} dz
\end{equation}
While $\mathcal{M}_h$ must technically be finite to ensure the Poisson process is properly defined, we can replace $\mathcal{M}_h=\infty$ in the upper limit of equation \ref{eqn_nlambda} without loss of precision.
Replacing $(\mathcal{M}_i, z_i)$ with $\bm{\theta}_i$, and expressing this in terms of the observed data $D$:
\begin{equation} \label{eqn_full}
    P(\bm{D} | \bm{\rho}) = \int d^N{\bm{\theta}} \, P(\bm{D} | \{\bm{\theta}_1, \dots, \bm{\theta}_N\}) \, P(\{\bm{\theta}_1, \dots, \bm{\theta}_N\} | \bm{\rho})
\end{equation}
We note that because the fits were performed independently to each object, the first term within the integral is
\begin{equation}
P(\bm{D} | \{\bm{\theta}_1, \dots, \bm{\theta}_N\})  =  \prod_i P(\bm{D}_i | \bm{\theta}_i)
\end{equation}
Replacing the second term with the expression for the inhomogeneous Poisson process from equation (\ref{eqn_poisson}), we obtain
\begin{equation}
P(\bm{D} | \bm{\rho}) = \mathcal{Z_\lambda} \, \int d^N{\bm{\theta}} \, \prod_i P(\bm{D}_i | \bm{\theta}_i) \lambda(\bm{\theta}_i | \bm{\rho})
\end{equation}
which can be simplified -- again exploiting the independence of the fits to each object -- to 
\begin{equation} \label{eqn_like}
    P(\bm{D} | \bm{\rho}) =  \mathcal{Z_\lambda} \, \prod_i \, \int \, d\bm{\theta}_i \, P(\bm{D}_i | \bm{\theta}_i) \, \lambda(\bm{\theta}_i | \bm{\rho})
\end{equation}

To incorporate uncertainties from the posterior $P(\bm{\theta}_i|\bm{D}_i)$ for the inferred parameters $\bm{\theta}_i$ for each galaxy, we need to marginalize over the unknown parameters:
\begin{equation}
    P(\bm{D}_i|\bm{\rho}) = \int P(\bm{D}_i | \mathcal{M}_i,z_i) \lambda(\mathcal{M}_i,z_i) d\bm{\theta}_i
\end{equation}
This represents the likelihood-weighted average of the probability of our continuity model over all possible values of $\bm{\theta}_i$.

We approximate this integral using a set of $m$ samples $\{\bm{\theta}_{i,1},\dots,\bm{\theta}_{i,m}\}$ drawn from the posterior $P(\bm{\theta}_i|\bm{D}_i)$ of each object. Assigning each sample an importance weight 
\begin{equation}
w_{i,j} = \frac{1}{P(\bm{\theta}_{i,j})}
\end{equation}
then allows us to approximate this integral as
\begin{equation}
    \int P(\bm{D}_i|\bm{\theta}_i) P(\bm{\theta}_i|\bm{\rho}) d\bm{\theta}_i \approx \frac{\sum_{j=1}^{m} w_{i,j} P(\bm{\theta}_{i,j}|\bm{\rho})}{\sum_{j=1}^{m} w_{i,j}}
\end{equation}
$P(\bm{\theta}_{i,j})$ are the chosen priors on mass and redshift during the SED fits performed by \prospector{}. The adopted redshift prior is a delta function while the stellar mass prior is uniform in logarithmic space. Given that this analysis also operates on $\mathcal{M}\equiv\log(M)$ rather than $M$, it follows that all $w_{i,j}$ are constant. In practice, we find that the results converge for $m \gtrsim 10$ posterior samples, and we take $m=50$.

Substituting our approximations and definitions into equation \ref{eqn_like}, our log-likelihood becomes
\begin{equation}
    \label{eq:poisson_lnprob}
    \boxed{\ln P(\bm{D} | \bm{\rho}) \approx \sum_{i=1}^N \ln \Big(\sum_{j=1}^m \lambda(\mathcal{M}_{i,j}, z_{i,j} | \bm{\rho})\Big) - N_\lambda(\lambda | \bm{\rho})}
\end{equation}
The subsequent section addresses the definition of the rate term $\lambda$.

\subsection{The Schechter Function}
\label{sec:schechter}
For our continuity model, the rate function can be evaluated as
\begin{equation} \label{eqn:rate}
\lambda(\mathcal{M},z) = \frac{\partial^2 N(\mathcal{M},z)}{\partial z \partial \mathcal{M}} = \Phi(\mathcal{M},z) V_{co}(z)
\end{equation}
where $V_{co}(z)$ is the differential comoving volume element and $\Phi(\mathcal{M},z)$ is the (un-normalized) stellar mass function evaluated at redshift $z$. This is an intuitive result: the occurrence rate of galaxies is proportional to their space density multiplied by the differential co-moving volume element, $V_{co}$.

We adopt the sum of two Schechter functions to describe the evolution of the mass function $\Phi(\mathcal{M},z)$. The logarithmic form of a single Schechter function is written as:
\begin{equation} \label{eqn:dblschech}
    \Phi(\mathcal{M}) =
    \ln(10)\phi_* 10^{(\mathcal{M}-\mathcal{M_*})(\alpha+1)}
    \exp{(-10^{\mathcal{M}-\mathcal{M_*}})}
\end{equation}
for a given $\phi_*$, $\mathcal{M}_* \equiv \log_{10}(M_*)$, and $\alpha$.
The integral of this function over mass from some lower limit $\mathcal{M}_c$ to infinity gives the expected total number density of galaxies $N_\Phi$:
\begin{equation}\label{eqn:schech_norm}
    N_\Phi \equiv \int_{\mathcal{M}_c}^{\infty} \Phi(\mathcal{M}) d\mathcal{M} = \phi_* \; \Gamma(\alpha+1,10^{\mathcal{M}_c-\mathcal{M}_*})
\end{equation}
with $\Gamma$ representing the upper incomplete gamma function. When necessary, this can be used to normalize the Schechter function such that it integrates to unity:
\begin{equation}
P(\mathcal{M}|\phi_*,\mathcal{M_*},\alpha,\mathcal{M}_c) = 
\begin{cases}
\frac{\Phi(\mathcal{M}|\phi_*,\mathcal{M}_*,\alpha)}{N_\Phi} & \mathcal{M} \geq \mathcal{M}_c \\
0 & \mathcal{M} < \mathcal{M}_c
\end{cases}
\end{equation}
We take $\mathcal{M}_*$ to be the same for both Schechter functions, as is standard fitting double Schechter functions (e.g., {Baldry} {et~al.} 2012; {Muzzin} {et~al.} 2013; {Tomczak} {et~al.} 2014). 

We altogether have five parameters to describe the mass function at a fixed redshift: $\phi_1$, $\phi_2$, $\mathcal{M}_*$, $\alpha_1$, and $\alpha_2$. We model the evolution of $\phi_1$, $\phi_2$, and $\mathcal{M_*}$ with a quadratic equation in redshift, such that
\begin{equation}
\rho_i(z) = a_{0,i} + a_{1,i}z+a_{2,i}z^2
\end{equation}
where $a_{j,i}$ are the continuity model parameters ({Drory} {et~al.} 2009; {Leja} {et~al.} 2015; {Wright} {et~al.} 2018). We fit redshift-independent values for $\alpha_1$ and $\alpha_2$ in order to limit degenerate solutions. This results in an $N_{\mathrm{dim}}=11$ model.

In practice, we do not fit directly for the quadratic coefficients but instead for the anchor points $\bm{\rho}(z_1)$, $\bm{\rho}(z_2)$, and $\bm{\rho}(z_3)$ from which the coefficients can then be derived. This is done because it is more straightforward to express physically meaningful priors on the anchor points. The redshifts for the anchor points are taken to be $z_{1} = 0.2$, $z_2=1.6$, and $z_{3} = 3$; these are chosen to bracket the redshift range of the surveys and the results do not depend on this choice. The adopted priors are uniform for each parameter and the ranges are shown in Table \ref{table:priors}. The different priors for the two low-mass slopes are chosen in order to keep the two Schechter functions distinct.

We note that, by definition, $P(\mathcal{M}|\phi_*,\mathcal{M}_*,\alpha,\mathcal{M}_c)$ returns a probability of zero for any mass sample below the lower limit $\mathcal{M}_c$. This is only relevant for objects whose posterior median masses are above the mass limit such that they enter the sample selection, but whose mass posteriors extend below the mass limit. We find that using more complex forms of the selection function has a negligible impact on our results.

\begin{figure*}[t!]
\begin{center}
\includegraphics[width=0.95\linewidth]{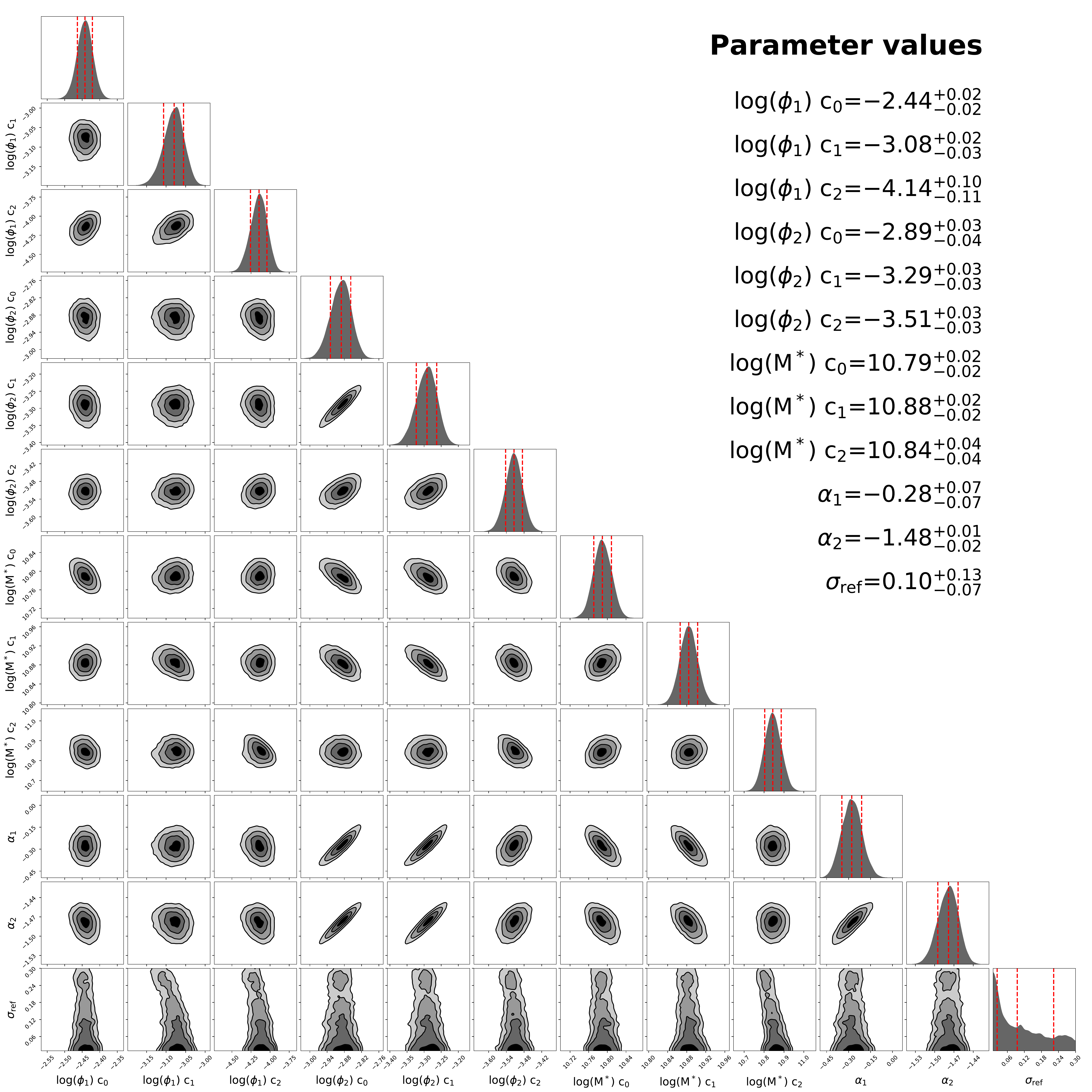}
\caption{Joint constraints for the continuity model fit. The diagonal panels show the 1D posterior for each model parameter. The off-diagonal panels show the 2D posterior for each pair of parameters. The 1D posterior median and 16$^{\mathrm{th}}$/84$^{\mathrm{th}}$ percentiles are indicated above the diagonal elements. The many parameter degeneracies highlight the utility of linking parameters between redshift bins. Appendix \ref{sec:guide} provides a guide to convert the continuity model parameters into the mass function at an arbitrary redshift $z_0$.}
\label{fig:corner}
\end{center}
\end{figure*}

\begin{deluxetable}{cc}
\tablecaption{Free parameters and priors of the continuity model. \label{table:priors}}
\tablehead{
\colhead{Parameter} & \colhead{Prior range}
}
\startdata
log($\phi_1$/Mpc$^{3}$/dex) & $-6,-2$ \\
log($\phi_2$/Mpc$^{3}$/dex) & $-6,-2$ \\
log(M$_*$/M$_{\odot}$) & $10,12$ \\
$\alpha_1$ & $-0.5,1$ \\
$\alpha_2$ & $-2,-0.5$ \\
log($\sigma_{\mathrm{ref}}$/dex) & $-2, -0.5$
\enddata
\end{deluxetable}

\subsection{Sample Variance}
\label{sec:cv}
In practice, the physics of structure formation causes galaxies to be distributed in clumps and voids. This means that a survey over a discrete volume can be subject to significant sample variance, sometimes referred to as cosmic variance. Accordingly, the mean density field $\lambda'(\mathcal{M},z)$ within any observed volume is likely to differ from the true mean density field $\lambda(\mathcal{M},z)$.

Since we are interested in inferring $\lambda(\mathcal{M},z)$ rather than $\lambda'(\mathcal{M},z)$, we wish to marginalize over this sampling variance:
\begin{equation}
    P(\lambda) \propto \int \left(e^{-N_{\lambda'}} \prod_{i=1}^N \lambda'_i \right) P(\lambda|\lambda') d\lambda'
\end{equation}
where $\lambda'_i \equiv \lambda'(\mathcal{M}_i,z_i)$.
Performing this integral is computationally challenging because we have to marginalize over $N$ objects for all possible values of $\lambda'(\mathcal{M},z)$, which in theory should be correlated in $\mathcal{M}$ and $z$.

We make two significant approximations in order to evaluate this integral. The first approximation is that the expected number of counts $N_{\lambda'}$ is roughly independent of any particular realization of the density field $\lambda'$. In other words, there are a sufficiently large number of objects such that the correlation between $N_{\lambda'}$ and realizations of the density field $\lambda'_i$ is small, and therefore the integral can be separated into two components:
\begin{equation}
    P(\lambda) \sim e^{-M_\lambda} \times \left[\int \left(\prod_{i=1}^N \lambda'_i \right) P(\lambda|\lambda') d\lambda'\right]
\end{equation}
where
\begin{equation}
    M_\lambda \equiv -\ln \int P(\lambda|\lambda') e^{-N_{\lambda'}} d\lambda'
\end{equation}
can be interpreted as the expected number of galaxies marginalizing over the unknown realizations of the density field $\lambda'(\mathcal{M},z)$ in the survey.

The second approximation we make is that the fluctuations in $\lambda'(\mathcal{M},z)$ constitute pure white noise such that errors in $\mathcal{M}$ and $z$ are independent of one another. In other words, the covariance between any two points $\mathcal{M},z$ and $\mathcal{M}^{'},z^{'}$ is zero, except in the case where the two points are exactly identical. While not strictly true, this approximation is needed to make the problem computationally feasible, as including spatial correlations would necessitate inverting very large ($\sim10^5 \times 10^5$) matrices. We take this white noise to be distributed as a Gaussian in logarithmic space with some amplitude $\sigma_{\rm samp}$:
\begin{equation}
\log(\lambda') \sim G(\log(\lambda), \sigma_{\mathrm{samp}})
\end{equation}
This allows us to factor the integral over objects into $N$ individual components, all of which can be evaluated independently:
\begin{equation}
    \int \left(\prod_{i=1}^N \lambda'_i \right) P(\lambda|\lambda') d\lambda' = \prod_{i=1}^{N} \int \lambda'_i \, P(\lambda_i|\lambda'_i) d\lambda'_i
\end{equation}
Each term in this integral is simply the expectation value of $P(\lambda|\lambda')$. Since the noisy rate $\lambda'$ is assumed to be log-Gaussian, this simply evaluates to:
\begin{equation}
    \Lambda_i \equiv \int \lambda'_i \, P(\lambda_i|\lambda'_i) d\lambda'_i = \exp(\log(\lambda_i) + \sigma_{\rm samp}^2/2)
\end{equation}
For the noiseless case where $\sigma_{\rm samp}=0$, this reduces to $\Lambda_i=\lambda_i$ as expected.

Altogether, this modifies the likelihood equation to be
\begin{equation}
    \label{eqn:full_likelihood}
    \boxed{\ln P(\bm{D} | \bm{\rho}) \approx \sum_{i=1}^N \ln \Big(\sum_{j=1}^m \Lambda(\mathcal{M}_{i,j}, z_{i,j} | \bm{\rho})\Big) - M_\lambda(\lambda | \bm{\rho})}
\end{equation}
The sampling variance term has the net effect of slightly increasing the model number density, implying that the observed mass function is slightly offset to higher number densities than the intrinsic mass function. This is a natural consequence of assuming log-normal density fluctuations: a small window into the density field is likely to be biased high. In the limit of a wide survey which includes both positive and negative fluctuations -- the sum of which will more accurately encompass the average density field -- the approximation is less appropriate. As will be seen later, this term does not introduce any significant bias in the derived mass function in this work.

The next step is writing down a functional form for $\sigma_{\mathrm{samp}}$. Typically, the uncertainty due to sampling variance is based on the geometry of the volume in which the mass function is inferred (e.g., {Driver} {et~al.} 2011). This is because sampling uncertainty is inherently a counting statistic: it encapsulates the distribution of possible differences between the mass function inferred in some volume and the `true' mass function. Here, rather than counting galaxies in a discrete volume, the continuity model infers a smooth evolution of a distribution function over a large volume. Accordingly, the sampling uncertainty $\sigma_{\mathrm{samp}}$ is instead modeled as an increase in the uncertainty of the number density for a {\it single} object.


Importantly, galaxy clustering and bias are functions of both stellar mass and redshift ({Moster} {et~al.} 2011), suggesting that the sampling variance term should have a stellar mass dependence. {Moster} {et~al.} (2011) calculates the expected size of sampling variance effects from models of dark matter evolution. They translate this to galaxies using a halo occupation distribution model ({Moster} {et~al.} 2010). We adopt the stellar mass dependence $\beta$ from equation (13) of that study, normalized such that an object with a stellar mass of 10$^{10}$ M$_{\odot}$ at the redshift midpoint of our sample, $z=1.6$, has $\beta=1$. The resulting expression for sampling variance is
\begin{equation} \label{eqn:cv}
\sigma_{\mathrm{samp}} = \ln\left(\frac{e^{\sigma_{\mathrm{ref}}}-1}{\beta(M_{stellar},z)} + 1\right)
\end{equation}
where $\sigma_{\mathrm{ref}}$ is the sampling uncertainty for a galaxy with a stellar mass of of 10$^{10}$ M$_{\odot}$ at $z=1.6$. The prior for $\sigma_{\mathrm{ref}}$ is uniform in logarithmic space (i.e., the Jeffreys prior) between 0.01 and 0.3 (implying that $\log\sigma_{\mathrm{ref}}$ goes from -2 to -0.5; see Table \ref{table:priors}). We show the results of fitting the continuity model to mock galaxies from a cosmological simulation in Appendix \ref{sec:mock_sigma}, demonstrating that the true value of $\sigma_{\mathrm{ref}}$ should fall in the region defined by the priors.

\section{Results}
\label{sec:results}
\begin{figure*}[t!]
\begin{center}
\includegraphics[width=0.8\linewidth]{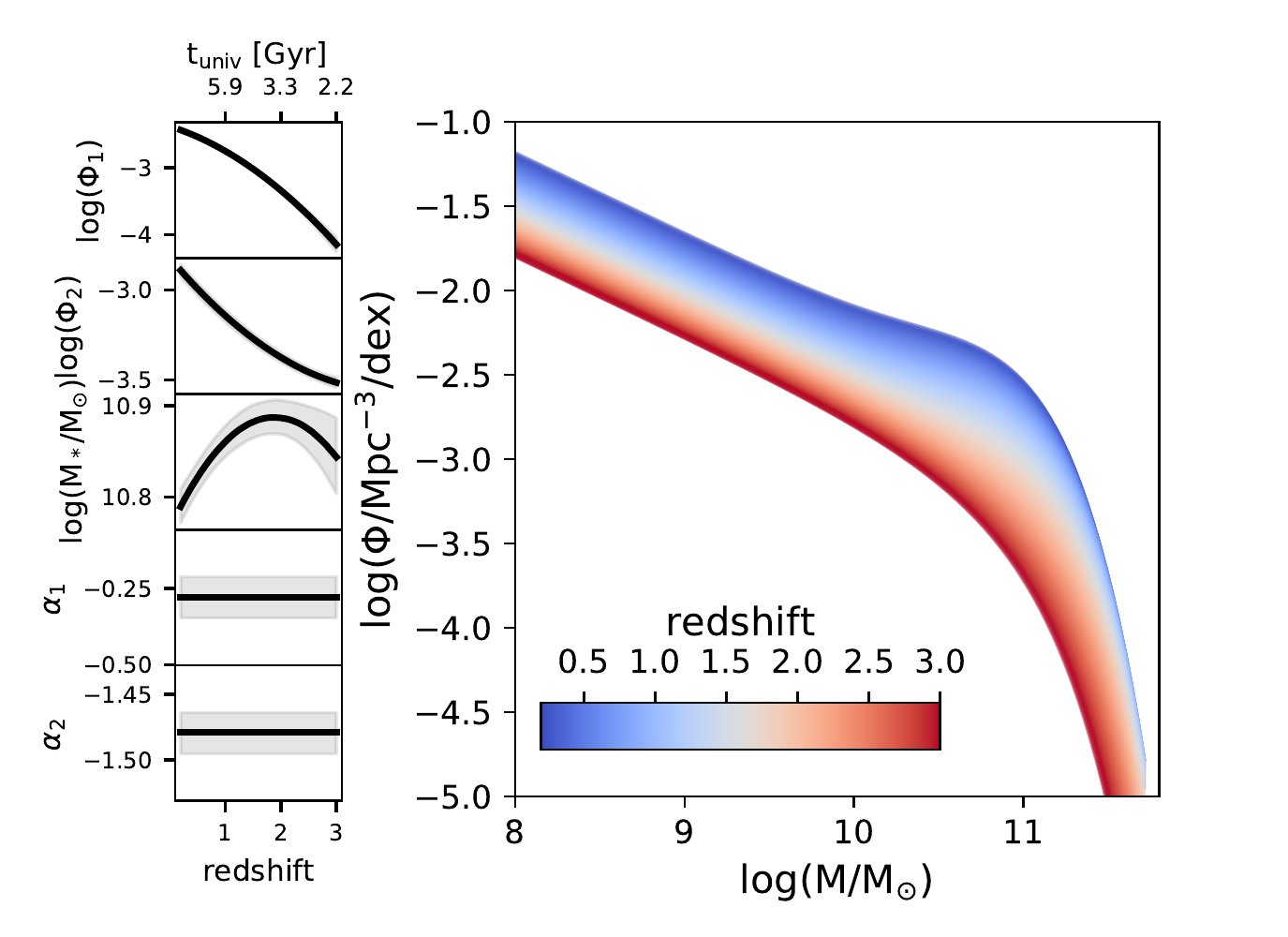}
\caption{The left panels show the redshift evolution of the double Schechter parameters while the right panel shows redshift evolution of the stellar mass function. The lines in the left panel are the median of the posterior, while the shaded regions indicate the $1\sigma$ range. The color shading in the right panel indicates the posterior median.}
\label{fig:mf_time_evolve}
\end{center}
\end{figure*}
The continuity model with the sample variance term included has 12 parameters, including 11 for the evolution of the mass function and 1 for the sample variance. The nested sampling code {\texttt dynesty} ({Speagle} 2020) is used to sample the model posteriors. Figure \ref{fig:corner} shows the model posteriors and covariances. There are multiple parameter degeneracies in this model, identifiable via diagonal shapes in the joint posterior panels. This underscores the utility of assuming continuity between redshift bins.

\subsection{The growth of the stellar mass function}
The continuity model parameters and their associated uncertainties are accessible in Figure \ref{fig:corner}. The redshift evolution of the Schechter parameters and the derived stellar mass functions are shown in Figure \ref{fig:mf_time_evolve}. Appendix \ref{sec:guide} provides a guide to convert the continuity model parameters into the mass function at an arbitrary redshift $z_0$, along with \texttt{python} code to perform this task.

Broadly speaking, the mass function grows as a function of cosmic time, consistent with many previous analyses in the literature. At high redshifts, the Schechter function with the steeper slope ($\phi_2$) dominates. As the massive end builds up with time, the more shallow component ($\phi_2$) begins to dominate. The exponential cutoff parameter, M$_*$, shows relatively little evolution with time, consistent with other analyses of the mass function (e.g., {Marchesini} {et~al.} 2009; {Muzzin} {et~al.} 2013). The massive end of the mass function is relatively stable after $z\sim0.8$. This is discussed further in Section \ref{sec:massive}.

By construction, the low-mass slopes have no redshift dependence. This is done to limit the degeneracies between the low-mass slope and other Schechter parameters. Above $z\gtrsim1.5$, the 3D-HST survey is not sufficiently deep to constrain all five Schechter parameters, and the polynomial prior typically induces significant time evolution once a degeneracy is encountered. The 11 parameters of the current mass function parameterization are sufficient to describe the observed number densities to the mass-complete limit of the surveys considered here. We emphasize that this modeling choice is unrelated to whether the low-mass slope {\it should} evolve with time. There do exist theoretical predictions that the low-mass slope should remain constant with time (e.g., {Kelson}, {Benson}, \& {Abramson} 2016), though generally in cosmological hydrodynamical simulations the low-mass slope becomes more shallow with time (e.g., {Pillepich} {et~al.} 2018; {Ma} {et~al.} 2018). This number density evolution will still be accurately reflected in the model posteriors, as presented here, assuming it is above the mass-complete limit of the survey.

\begin{figure*}[t!]
\begin{center}
\includegraphics[width=0.95\linewidth]{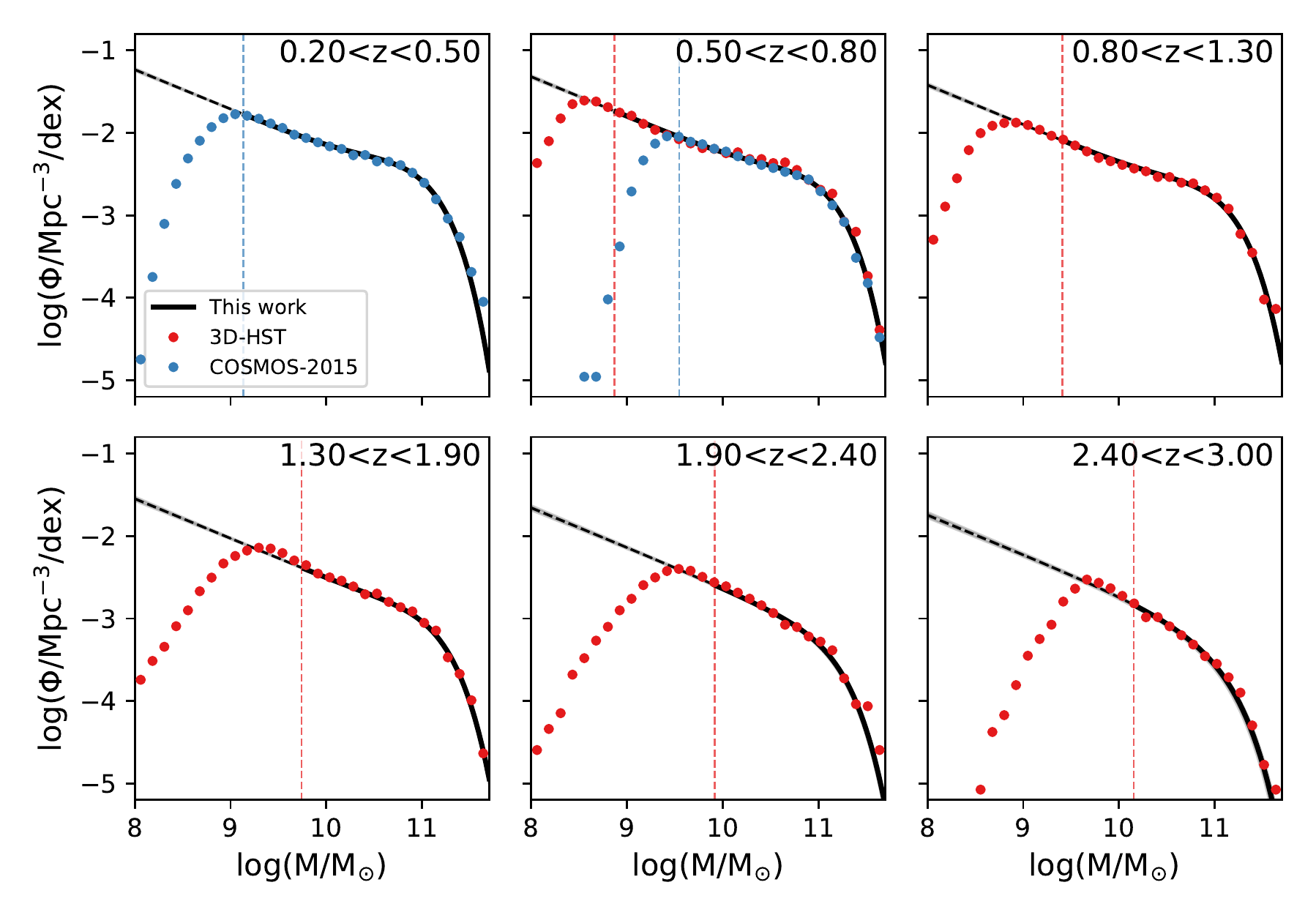}
\caption{Comparing the continuity model posteriors to the galaxy counts from the 1/V$_{max}$ method. This demonstrates that the continuity model accurately reproduces the binned mass functions (though it was not fit to them). The dotted lines indicates the mass completeness limit calculated at the upper limit of the redshift bin. Galaxies from COSMOS-2015 and 3D-HST are plotted separately as red and blue symbols; notably, there is consistency in the region where these surveys overlap. The light grey shading indicates the 1$\sigma$ uncertainty in the mass function posterior -- this is thinner than the width of the line in most cases.}
\label{fig:compare_to_bins}
\end{center}
\end{figure*}

Figure \ref{fig:compare_to_bins} compares the mass function inferred with the continuity model to the number density estimates from the $1/V_{\mathrm{max}}$ method. There is good agreement, demonstrating that the continuity model (Section \ref{sec:mf_model}) is sufficiently flexible to describe the growth of the mass function over the $10$ Gyr covered in this redshift range. The slight offset of the continuity model to lower number densities at high masses and redshifts is caused by both the larger stellar mass uncertainties and by the increased sampling uncertainty in this regime. There is no hint of a decrease in number counts near the adopted stellar mass limit, implying that the derived mass-complete limits are an acceptable description. In fact, they seem to be a conservative choice, as the model continues to accurately describe the binned galaxy counts $\sim$0$.2-0.3$ dex below the adopted mass completeness limits.

\subsection{Comparison to the standard technique}
\label{sec:ml_comp}
We contrast the results of our continuity model with a Schechter function fit to the 1/V$_{max}$ points in fixed redshift bins. The mass function is divided into eight redshift bins, detailed in Table \ref{table:zbins_ml}.

\begin{deluxetable}{c}
\tablecaption{Redshift bins adopted for the Schechter fits to the 1/V$_{max}$ estimates \label{table:zbins_ml}}
\tablehead{
\colhead{adopted redshift bins} }
\startdata
(0.2, 0.5) \\
(0.5, 0.8) \\
(0.8, 1.1) \\
(1.1, 1.4) \\
(1.4, 1.8) \\
(1.8, 2.2) \\
(2.2, 2.6) \\
(2.6, 3.0)
\enddata
\end{deluxetable}

The uncertainties are taken as the quadratic sum of Poisson uncertainties and sampling variance uncertainties. The sampling variance uncertainties are generated with the {Driver} {et~al.} (2011) cosmic variance calculator. The fit is performed with \texttt{emcee} ({Foreman-Mackey} {et~al.} 2013), with walker convergence assessed by eye after two burn-in phases and 10,000 iterations. The parameter priors are set to match the associated anchor point priors from the continuity model (Table \ref{table:priors}). During the fit, the intrinsic mass function is convolved with a Gaussian of width $\sigma=0.05$ dex to simulate the net effect of stellar mass uncertainties.

\begin{figure*}[t!]
\begin{center}
\includegraphics[width=0.95\linewidth]{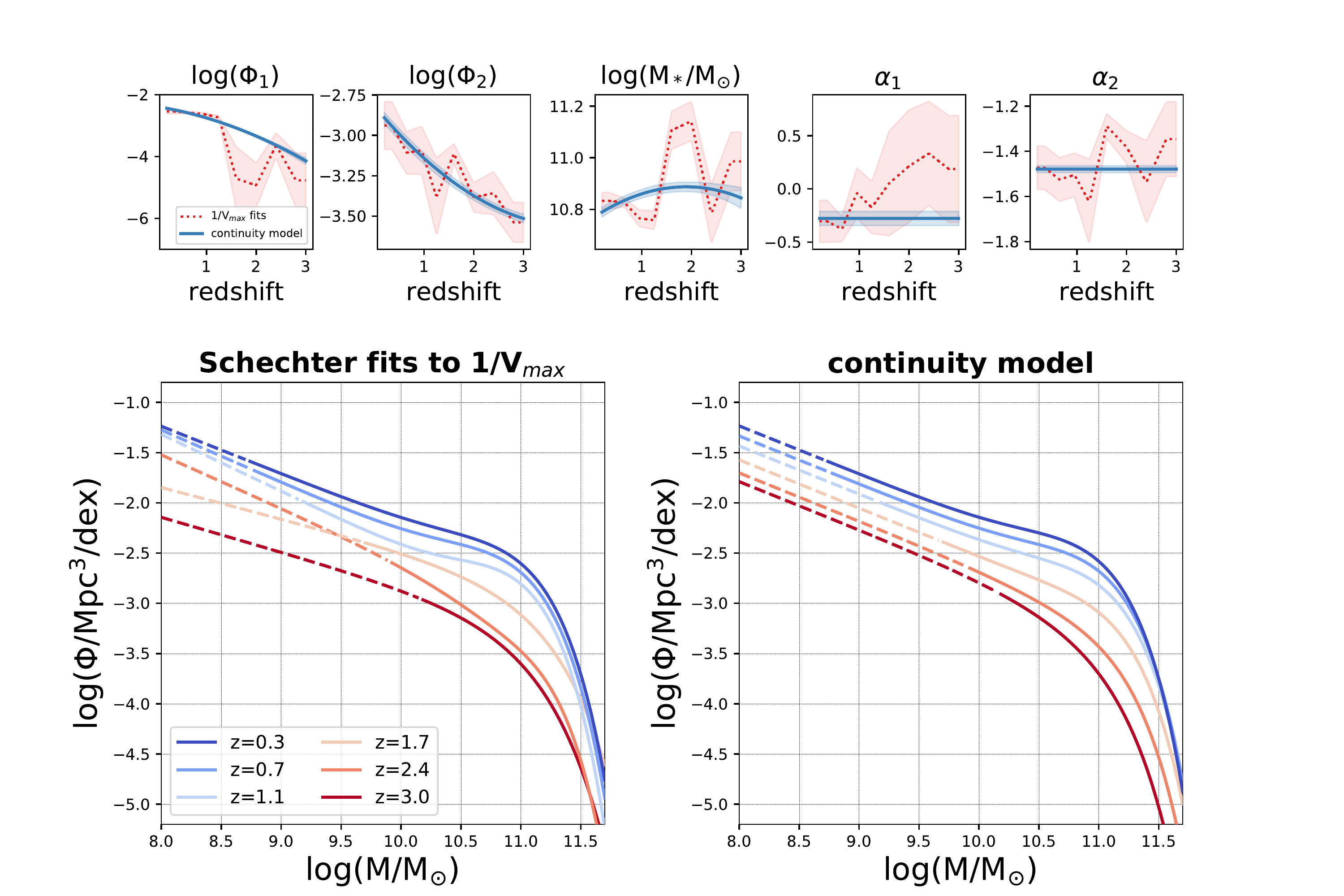}
\caption{Comparing the continuity model posteriors to a standard Schechter fit to the 1/V$_{max}$ estimates. The top panels show the redshift evolution of the Schechter parameters while the bottom panels show the redshift evolution of the mass function. The lines indicate the posterior median values; the shaded regions indicate $1\sigma$ uncertainties. For clarity, no uncertainties are shown for the mass functions. The dotted lines indicates an extrapolation below the mass-incomplete limit. The posterior differences come from structural differences in the two techniques: the 1/V$_{max}$ fits describe the stellar mass function in only a subset of the survey volume, while the continuity model is simultaneously constrained by the mass and redshift of every galaxy in the survey.}
\label{fig:compare_classic_fit}
\end{center}
\end{figure*}

\begin{figure*}
\begin{center}
\includegraphics[width=0.95\linewidth]{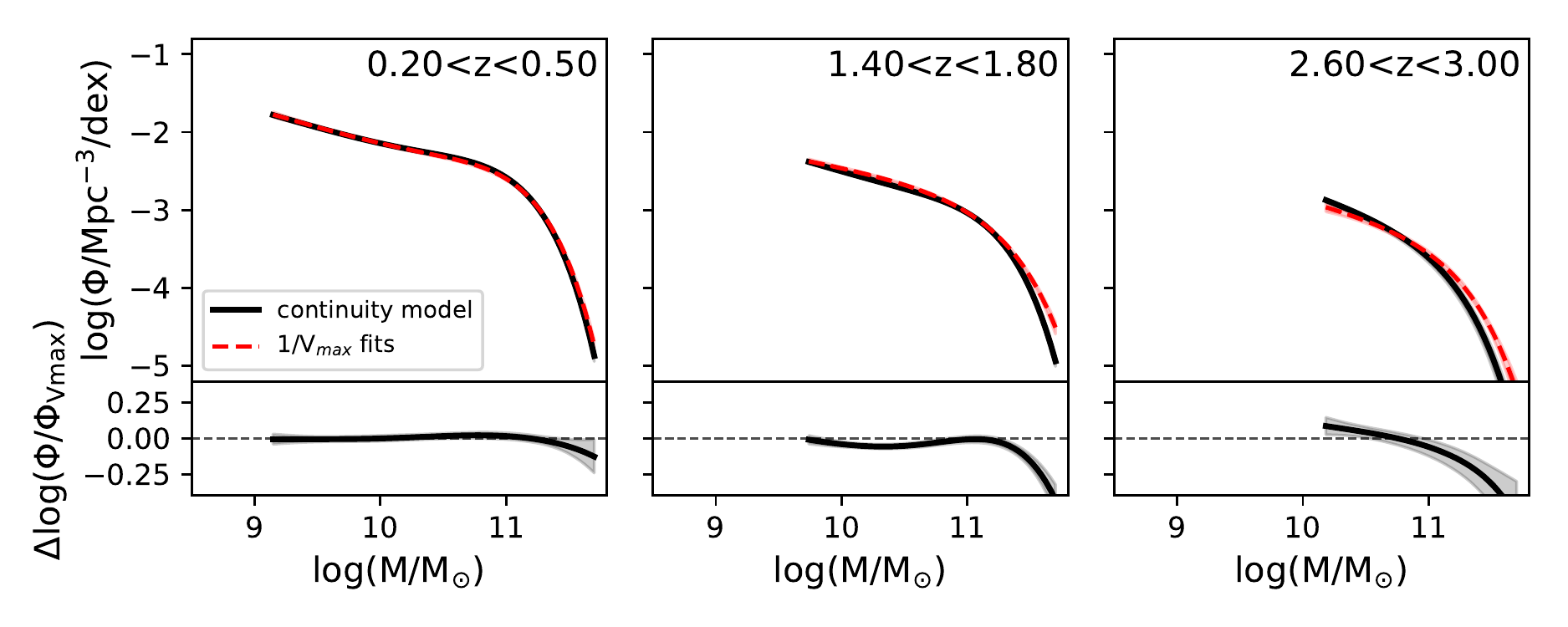}
\caption{Comparing the mass function inferred with the continuity model with the fit to the 1/V$_{max}$ points. The comparison is limited to masses above the mass-completeness limit. There is generally good agreement, though there are deviations at high masses and near the low-mass limit for reasons discussed in Section \ref{sec:ml_comp}. The upper panels show the mass functions while the lower panels highlight the residuals between the two fits. The lines are the posterior median and the shaded regions indicate the 1$\sigma$ uncertainty. The 1/V$_{\max}$ points are estimated inside the indicated redshift interval while the continuity model predictions are generated at the midpoint of the redshift interval.}
\label{fig:compare_vmax}
\end{center}
\end{figure*}

Figure \ref{fig:compare_classic_fit} contrasts the redshift evolution of the stellar mass functions derived from fitting the 1/V$_{max}$ points and from the continuity model. Figure \ref{fig:compare_vmax} compares the mass functions directly and highlights the residuals. Broadly the two approaches produce similar mass functions, but there are differences in detail. First, the uncertainty contours for the continuity model are much smaller. This occurs because the continuity model is more constrained than the standard approach. This is expected: the standard technique aims to describe the mass function in a specific volume, whereas the continuity model is additionally constrained by the mass functions in the entire survey volume. The continuity model effectively requires that the underlying galaxy populations are continuous in time.

Second, while there is generally good agreement between the two approaches above the stellar mass limit, there are differences in the extrapolation of the mass functions down to log(M/M$_{\odot})=8$. By fixing the faint-end slopes, the continuity model ensures that the extrapolation to lower masses is stable with redshift. In the standard approach, this extrapolation is more sensitive to variations in galaxy counts within the fixed volume. The continuity of the extrapolated fits is helpful in ensuring consistent results when using mass functions as inputs to models (e.g., {Drory} {et~al.} 2009; {Weinmann} {et~al.} 2012; {Leja} {et~al.} 2015; {Tomczak} {et~al.} 2016; {Behroozi} {et~al.} 2019). 

Finally, the implied evolution of the massive end of the stellar mass function is different between the two techniques. The 1/$V_{max}$ fits suggest non-monotonic evolution, including negative evolution from $z=2$ to $z=1$ which reverses to growth from $z=1$ to $z=0.2$. In contrast, the continuity model infers a smooth build-up of stellar mass at higher redshifts and very little evolution below $z\sim1$. The immediate cause of this difference is the fact that the continuity model is constrained by the entire redshift evolution of the massive end rather than the counts in any specific volume. The ultimate cause of this difference is the different sensitivities to sampling variance between the two techniques. Massive galaxies are more strongly affected by sampling variance due to both their low intrinsic numbers and their high bias relative to the underlying density field (Section \ref{sec:cv}). This results in large number density uncertainties on the massive end in the 1/$V_{max}$ fit: taken at face value, this suggests that massive galaxies grow in a mixture of rapid bursts and mass-loss events. In addition to this, the continuity model direct adjusts for the large bias of massive galaxies at high redshift, slightly decreasing their inferred number densities. The evolution of the massive end (or lack thereof) is discussed further in Section \ref{sec:massive}.

\subsection{Stellar mass uncertainties}
\label{sec:mass_errors}
\begin{figure}[h!]
\begin{center}
\includegraphics[width=0.95\linewidth]{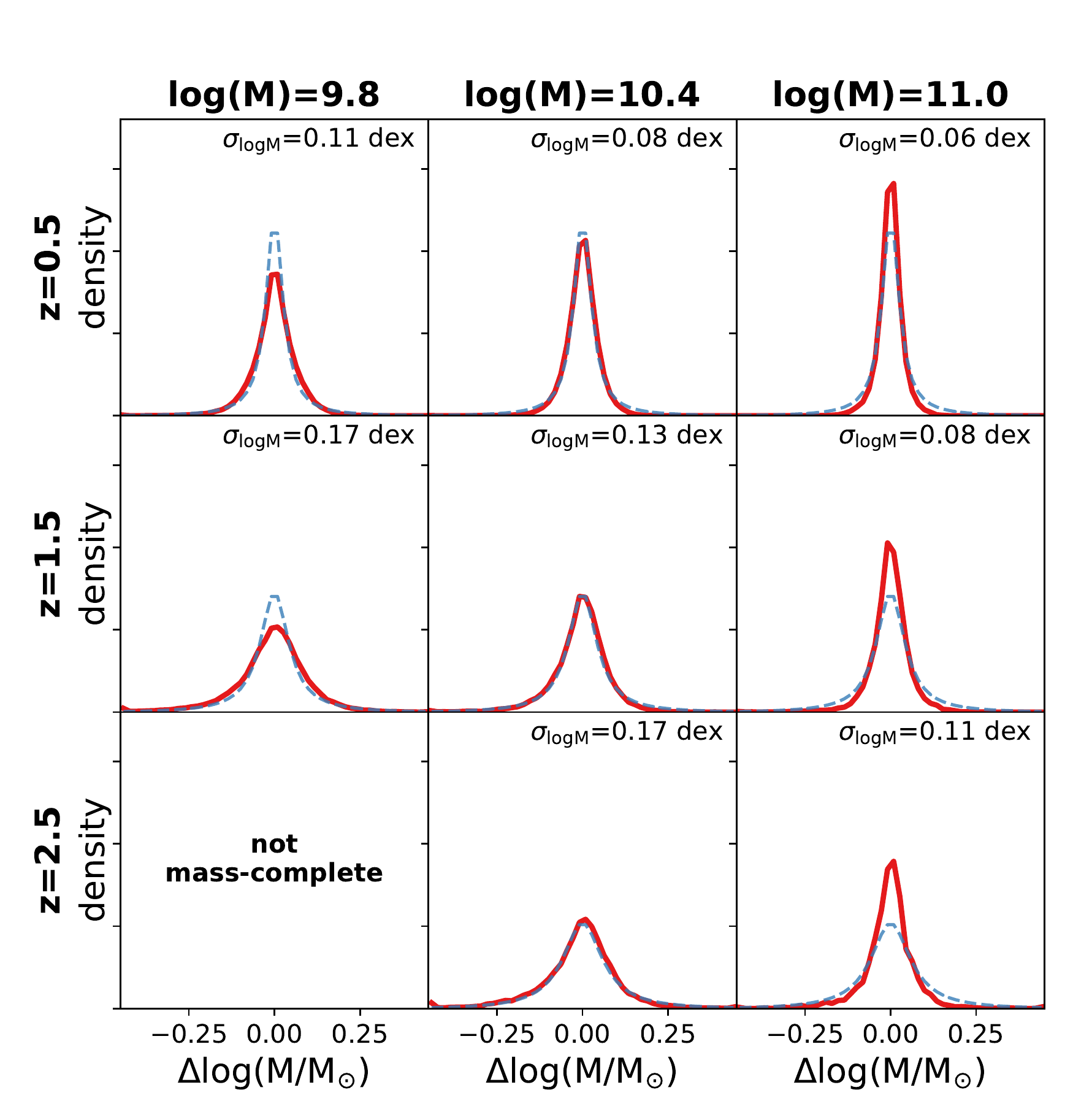}
\caption{The typical stellar mass uncertainty as a function of stellar mass (columns) and redshift (rows). An accurate inference of the mass function depends on accurate uncertainties, as the uncertainties are used to forward-model the observed stellar mass function. The uncertainty increases with decreasing stellar mass and increases with redshift. The red line illustrates the \prospector{} stellar mass uncertainties, compared with the {Davidzon} {et~al.} (2017) mass uncertainties indicated with the blue line in the top three rows. The 1$\sigma$ range is indicated in each panel. }
\label{fig:mass_uncertainty}
\end{center}
\end{figure}

Stellar mass uncertainties are interesting to examine in detail, as they are a key ingredient in forward-modelling the stellar mass function -- Appendix \ref{sec:mock} illustrates how the continuity model adjusts for the effect of stellar mass uncertainty. Specifically, accurate uncertainty estimates produce accurate corrections for the Eddington bias (discussed in Section \ref{sec:overview}) Figure \ref{fig:mass_uncertainty} shows the centered, summed stellar mass posteriors as a function of stellar mass and redshift, calculated in bins of $\delta z = 0.15$ and $ \delta \log \mathrm{M} = 0.15$. The stellar mass uncertainty ranges between $\sigma_{\mathrm{logM}} = 0.06-0.17$ dex. It increases at lower masses and at higher redshifts, consistent with the uncertainties on the observed fluxes. 

We also include the uncertainties from the most recent COSMOS stellar mass function ({Davidzon} {et~al.} 2017) in Figure \ref{fig:mass_uncertainty} for reference. These uncertainties include the effect of redshift uncertainty, whereas the \prospector{} uncertainties are calculated at a fixed redshift. Despite this, the \prospector{} mass uncertainties are comparable in size to {Davidzon} {et~al.} (2017). This suggests \prospector{} infers larger uncertainties on the stellar mass at fixed redshift, consistent with the greater flexibility of the \prospector{} model. The comparison also shows relatively lower \prospector{} uncertainties for massive objects, suggesting that photometric redshift uncertainties may contribute relatively more to the total uncertainty in massive objects.

Previous work has found that redshift uncertainty can be a significant component of the total stellar mass uncertainty, particularly at $z>2$ (e.g., {Caputi} {et~al.} 2011; {Grazian} {et~al.} 2015). {Grazian} {et~al.} (2015) compares photometric redshifts estimated with different codes, showing that the resulting dispersion in the inferred stellar mass functions is roughly the size of the Poisson error bars. However, these works focus on the high-redshift universe where photometric redshifts are the most uncertain, $\sigma_z \approx 0.05 (1+z)$. For $z>0.8$ the redshifts in this work are entirely from the 3D-HST survey, which utilizes space-based grism spectroscopy to infer more accurate redshifts with a measured accuracy of $\sigma_z = 0.02 (1+z)$ ({Bezanson} {et~al.} 2016). The scatter between spectroscopic redshifts and photometric/grism redshifts is minimized for brighter objects and for more massive objects ({Bezanson} {et~al.} 2016), suggesting that the redshifts of massive objects are relatively more well-determined, though the fraction of catastrophic outliers increases slightly to $\sim5\%$. The COSMOS redshifts have an even higher accuracy of $\sigma_z \approx 0.007 (1+z)$ ({Laigle} {et~al.} 2016).

While the grism-based redshifts are highly accurate and, as judged from the comparison to {Davidzon} {et~al.} (2017), likely do not dominate the stellar mass error budget, it is straightforward to generalize the methodology presented here to include redshift uncertainties. This will include the covariance between mass and redshift as well, rather than simply inflating the mass uncertainties at a fixed redshift by marginalizing over the redshift uncertainty. The limiting factor for including redshift uncertainties here is computational resources, but this will likely be alleviated in the future work with the use of techniques such as neural net emulation ({Alsing} {et~al.} 2019).

\subsection{Inferred sampling variance}
\begin{figure}[h!]
\begin{center}
\includegraphics[width=0.95\linewidth]{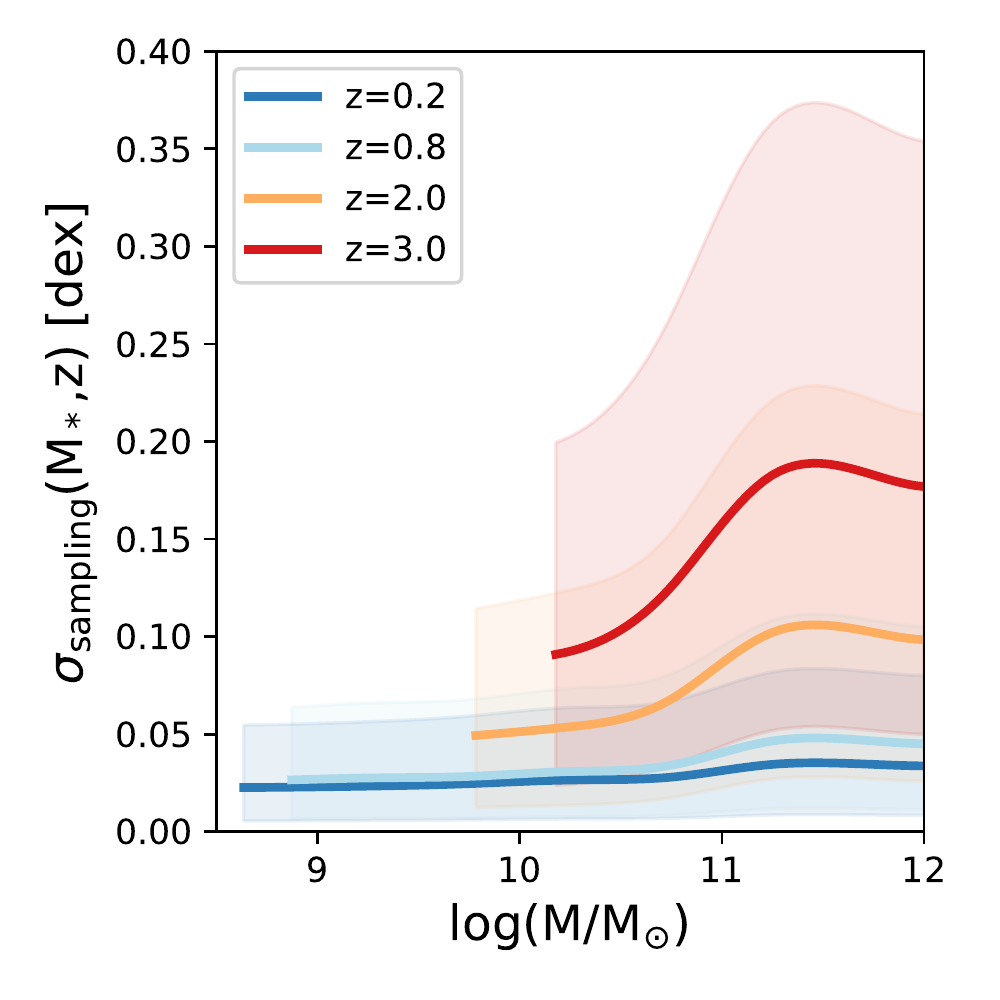}
\caption{The uncertainty in the mass function resulting from sampling variance. The sampling variance uncertainty is a free parameter in the continuity model constrained by the observed distribution of masses and redshifts. The lines show the median of the posterior while the shaded regions indicate the $1\sigma$ range. Notably, this should not be compared to the derived uncertainties in the mass function. Rather, it is the uncertainty included in the likelihood function for the {\it local} mass function of each galaxy.}
\label{fig:cv}
\end{center}
\end{figure}

Figure \ref{fig:cv} shows the model posteriors for the sampling variance term $\sigma_{\rm samp}$ from equation \eqref{eqn:cv}. The uncertainties are higher at higher redshifts and masses; this is driven by the bias of the underlying matter density field taken from {Moster} {et~al.} (2011). The posterior for the sampling uncertainty term is largely set by the chosen logarithmic prior (Figure \ref{fig:corner}). The stellar mass function uncertainty stemming from sampling variance is similar in magnitude to the sampling deviations in typical galaxy survey volumes (e.g., {Driver} {et~al.} 2011). The value derived observationally is also similar to the value derived by fitting the same model to galaxies in a cosmological simulation (Appendix \ref{sec:mock_sigma}).

The exact value of the added sampling variance term has little impact on the results presented in this work. This can be seen directly in the relatively small covariance between $\sigma_{\mathrm{sampling}}$ and the mass function parameters in Figure \ref{fig:corner}. The term with the greatest covariance is $\mathcal{M_*}$ at high redshift. This is unsurprising, as the sampling uncertainty term is maximized for high-redshift massive galaxies.

Even though the sampling uncertainties on the mass function for any specific object can be substantial (up to a factor of two), the continuity model infers the {\it mean} redshift evolution of the galaxy population. The constraints on the global mass function are thus expected to be stronger than the injected uncertainty due to sampling variance.

\subsection{The evolution of the total stellar mass density}
\label{sec:smd}
The total stellar mass density can be derived by integrating the Schechter functions down to some lower limit $\mathcal{M}_c\equiv\log(\mcomp)$. This produces
\begin{equation} \label{eqn:smd}
\rho_*(\mathcal{M}_c) = \phi_* 10^{\mathcal{M}_*} \; \Gamma(\alpha+2,10^{\mathcal{M}_c-\mathcal{M}_*})
\end{equation}
where $\Gamma$ is the upper incomplete gamma function and $\alpha$, $\phi_*$, and $\mathcal{M}_*\equiv\log_{10}(M_*)$ are the corresponding Schechter parameters.

\begin{figure*}[t]
\begin{center}
\includegraphics[width=0.95\linewidth]{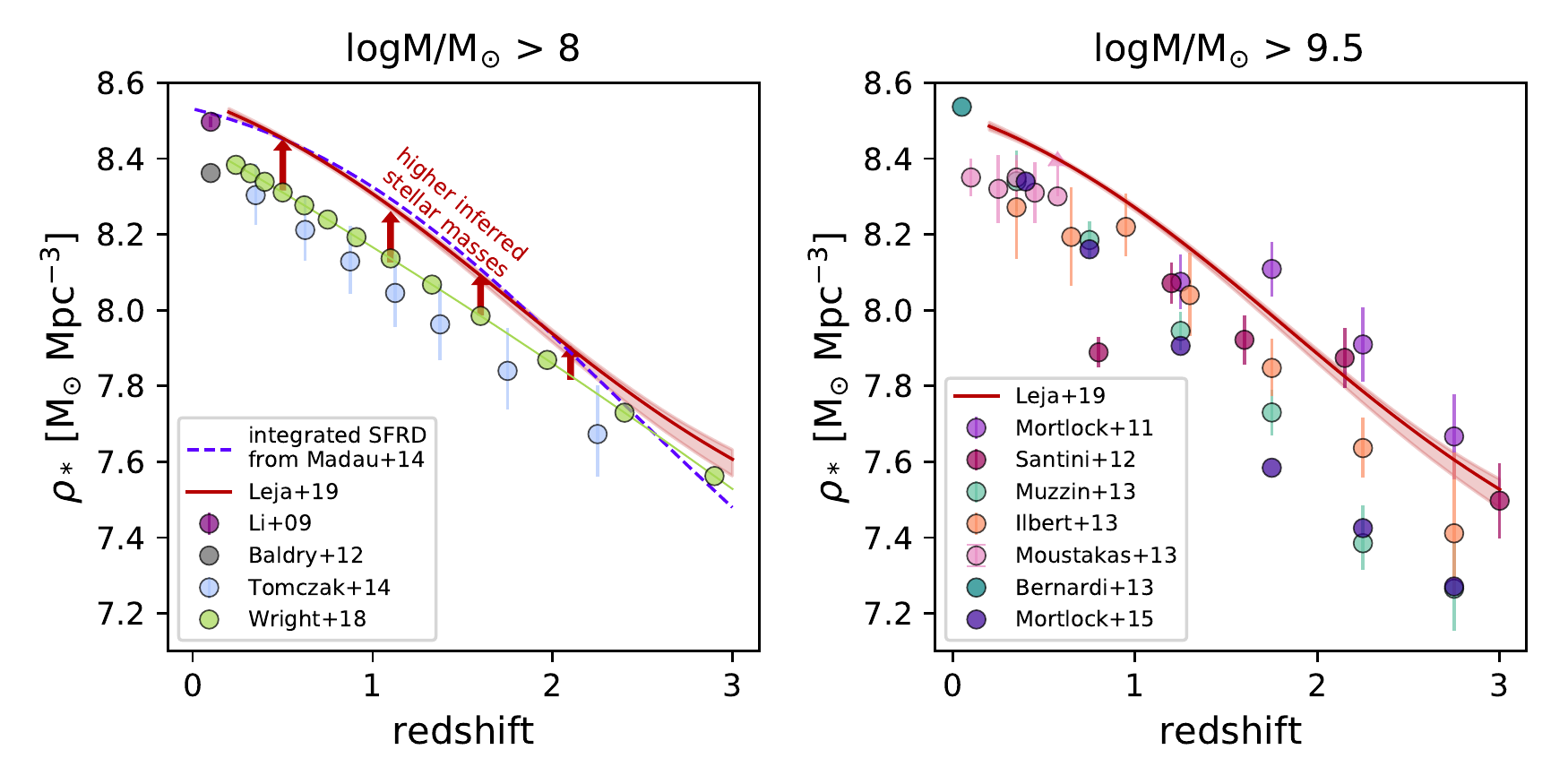}
\caption{Comparing the redshift evolution of the stellar mass density to literature measurements. This work infers $\sim$50\% higher stellar mass densities at $z>0.5$ than other measurements, with the rate of change maximized around $z\sim1.5$. This is largely due to differences in SED-fitting assumptions. The exception at $z\sim2$ ({Santini} {et~al.} 2012) focuses on a small field which likely has a relative overabundance of low-mass objects (see discussion in \ref{sec:smd}). The {Madau} \& {Dickinson} (2014) line is the integral of the star formation rate density, and the relative normalization at $z>0$ is highly sensitive to the adopted local stellar mass density. All indicated uncertainties are $1\sigma$.}
\label{fig:smd}
\end{center}
\end{figure*}

The redshift evolution of the total stellar mass density is shown in Figure \ref{fig:smd}. The build-up of the integrated stellar mass from $z=3$ to $z=0.2$ is relatively steady, with the derivative peaking around $z\sim1.5$. Notably, the location of this peak is lower than the peak of $z\approx1.9$ found in the consensus model of {Madau} \& {Dickinson} (2014).

Other measurements from the literature are included after converting their results to {Chabrier} (2003) initial mass functions ({Li} \& {White} 2009; {Baldry} {et~al.} 2012; {Bernardi} {et~al.} 2013; {Santini} {et~al.} 2012; {Moustakas} {et~al.} 2013; {Muzzin} {et~al.} 2013; {Tomczak} {et~al.} 2014; {Mortlock} {et~al.} 2015; {Davidzon} {et~al.} 2017; {Wright} {et~al.} 2018). For {Bernardi} {et~al.} (2013) we adopt the fits to the S\'{e}rsic light profile rather than the aperture photometry. For {Santini} {et~al.} (2012), we take the results of the {Bruzual} \& {Charlot} (2003) stellar populations fits (their Table 1). We also include the predictions from the integral of the {Madau} \& {Dickinson} (2014) star formation history, using the $z=0$ stellar mass density from {Gallazzi} {et~al.} (2008) and a canonical return {Salpeter} (1955) return fraction of $R=0.27$. We caution that this calculation is {\it highly} sensitive to the $z=0$ stellar mass density, such that a change of 0.02 dex in either direction produces dramatically different results. The stellar mass densities are integrated down to either $10^8$ or $10^{9.5}$ M$_{\odot}$, depending on the survey depths and data available in the literature.

The analysis presented here finds a systematically higher total stellar mass density at almost all redshifts than previous studies. This difference is largely due to differences in SED modeling, discussed further in Section \ref{sec:comparison}. One  exception is {Santini} {et~al.} (2012), who find a very high stellar mass density at $z\sim2$. This study uses data from early HST imaging of GOODS-S, and the high stellar mass density is driven by an unusually steep faint-end slope. We confirm through direct re-analysis that GOODS-S shows a steep faint-end slope compared to the other extragalactic fields. This faint-end slope is in contrast with subsequent deeper ({Tomczak} {et~al.} 2014) and wider ({Muzzin} {et~al.} 2013; {Mortlock} {et~al.} 2015) studies of extragalactic fields, and is also found in other studies of the GOODS-S field ({Mortlock} {et~al.} 2011). We conclude that the high stellar mass density in {Santini} {et~al.} (2012) may be driven by an overdensity of low-mass galaxies in the GOODS-S field. 

\section{Discussion}
\label{sec:discussion}
In this section we briefly compare the results of this study with other similar studies in the literature, and discuss the evolution (or lack thereof) in the massive end of the mass function.
\subsection{Comparison to previous mass functions in the literature}
\label{sec:comparison}
\begin{figure*}[t]
\begin{center}
\includegraphics[width=0.95\linewidth]{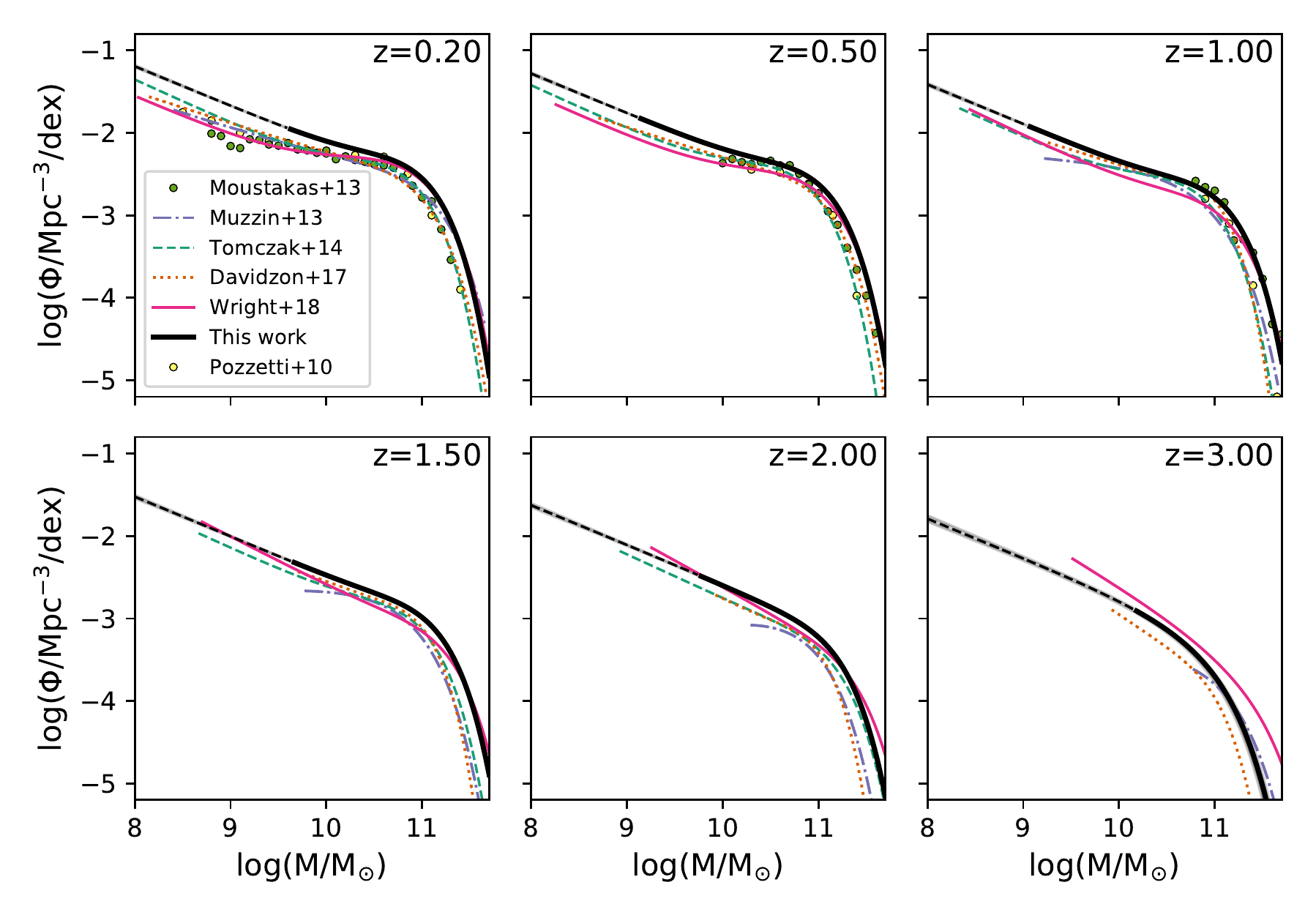}
\caption{Comparing to other mass function measurements in the literature. The difference is largely due to the fact that \prospector{} infers more massive galaxies than standard approaches in the literature. The offset peaks around $z\sim1$, an epoch where galaxies still had high sSFRs yet also had accumulated a substantial population of old stars.}
\label{fig:literature_comp}
\end{center}
\end{figure*}
Figure \ref{fig:literature_comp} compares the mass function from this work to comparable mass functions from the literature ({Pozzetti} {et~al.} 2010; {Moustakas} {et~al.} 2013; {Muzzin} {et~al.} 2013; {Tomczak} {et~al.} 2014; {Davidzon} {et~al.} 2017; {Wright} {et~al.} 2018). When needed, we interpolate the published mass function parameters between redshift bins. In a few cases the fitting technique is changed between redshift bins; in these cases, the interpolations fail and are not used. We take the double Schechter fits from {Muzzin} {et~al.} (2013) where available, otherwise the single Schechter fits with the free faint-end slope are used. The smooth parameterization of the {Tomczak} {et~al.} (2014) mass functions from {Leja} {et~al.} (2015) are used. We note that while {Wright} {et~al.} (2018) is based on the same surveys analyzed in this work, they use a photometric reduction pipeline which introduces systematic, wavelength-dependent offsets in the photometry of up to 0.15 magnitudes ({Andrews} {et~al.} 2017). The literature mass functions are truncated at their mass-complete limits, while the mass function from this work is extrapolated for comparison.

At almost all redshifts and masses, the continuity model in this work infers a higher number density at fixed stellar mass than other studies. This directly leads to the $30-100\%$ higher integrated stellar mass densities in Figure \ref{fig:smd}. Many of these mass functions were derived using either the same surveys analyzed in this work, or subsets of the same surveys. This overlap makes this comparison particularly interesting: these works are subject to the same redshift uncertainties, photometric uncertainties, and sampling uncertainties due to cosmic variance. {Pozzetti} {et~al.} (2010) is notable for using spectroscopic redshifts; the agreement between the {Pozzetti} {et~al.} (2010) mass function and the other mass functions, which primarily rely on photometric redshifts \footnote{Aside from {Moustakas} {et~al.} 2013, which uses prism redshifts}, is encouraging. This suggests that redshift uncertainties do not strongly affect the shape or normalization of the mass function, at least for $z<1$.

The new fitting methodology is not responsible for this difference; while the fitting methodology affects the extrapolation to lower masses, the size of the uncertainties, and the evolution of the massive end (see Section \ref{sec:ml_comp}), it does not cause the systematically higher stellar masses.

These higher masses originate from differences in the SED-fitting routines: \prospector{} infers systematically more massive galaxies than standard SED-fitting approaches, with the difference maximized at low masses. The causes of these differences are discussed in detail in {Leja} {et~al.} (2019b). The primary cause is the nonparametric SFHs used in \prospector{}. This approach produces mass-weighted ages that are $\sim3-5$ times older than standard parametric models ({Carnall} {et~al.} 2019; {Leja} {et~al.} 2019a), which in turn result in larger mass-to-light ratios and larger stellar masses. A second factor is that \prospector{} uses the FSPS stellar populations synthesis code, which infers $\sim0.05$ dex systematically larger masses than codes such as {Bruzual} \& {Charlot} (2003).

There are several pieces of independent evidence which support these elevated stellar masses. {Leja} {et~al.} (2019b) show that star formation histories inferred with standard SED-fitting approaches are far too short to be consistent with the build-up of the observed mass function, especially at low masses (see also {Wuyts} {et~al.} 2011). In contrast, the more extended star formation histories inferred with \prospector{} are in relatively good agreement with the observed evolution of the stellar mass function. {Leja} {et~al.} (2019b) further verifies that the higher stellar masses remain below the measured dynamical masses ({Bezanson}, {Franx}, \& {van  Dokkum} 2015). Importantly, these larger stellar masses increase the derivative of the stellar mass density by $\sim$0.2 dex, bringing it into agreement with the observed star formation rate density. However, while the systematically higher stellar masses from \prospector{} appear to resolve several issues with the standard approach, the overall consistency of this new picture of galaxy evolution must still be thoroughly tested. Key future tests include a more detailed comparison to dynamical masses (e.g. {Price} {et~al.} 2019), spectroscopic ages (e.g. {Belli}, {Newman}, \& {Ellis} 2019), comparison to spatially-resolved SED fits (e.g. {Sorba} \& {Sawicki} 2018), and verification of the SED-fitting methodology using realistic simulated SFHs (e.g. {Simha} {et~al.} 2014).

\subsection{The growth of massive galaxies since $z\sim1$}
\label{sec:massive}
There is an open question in the literature as to what extent the massive end of the stellar mass function (M$_*$ $\gtrsim10^{11.2}$ M$_{\odot}$) grows between $0<z<1$. While the most massive galaxies have little ongoing star formation, there are theoretical expectations for significant growth in the massive galaxies at late times due to galaxy-galaxy mergers ({De Lucia} {et~al.} 2006; {De Lucia} \& {Blaizot} 2007; {Naab}, {Johansson}, \& {Ostriker} 2009; {Behroozi} {et~al.} 2013; {Tacchella} {et~al.} 2019). There is also observational evidence that the massive galaxies experience substantial growth through mergers, including the observed size evolution of massive galaxies since $z\sim2$ ({van Dokkum} 2008; {Bezanson} {et~al.} 2009; {van Dokkum} {et~al.} 2010; {Belli} {et~al.} 2014; {Mowla} {et~al.} 2019) and the lower metallicities and younger ages observed on the outskirts of nearby massive galaxies ({Rowlands} {et~al.} 2018; {Oyarz{\'u}n} {et~al.} 2019). 

Despite these expectations, {Moustakas} {et~al.} (2013) constrain the evolution of the mass function over a wide area of $\sim$5.5 degrees$^2$ and find zero net evolution in the massive end of the mass function since $z=1$. Similarly, {Bundy} {et~al.} (2017) studies 139 deg$^2$ in the Stripe 82 Massive Galaxy Project and finds no evolution in the massive end of the mass function over $0.3< z< 0.65$. These results are consistent with the continuity model presented here, which also finds very little observed growth of the massive end of the mass function since $z\sim0.8$ (e.g, Figure \ref{fig:mf_time_evolve}). Number density arguments suggest that a stellar mass function which is constant in time also implies negligible mass evolution in individual objects ({van Dokkum} {et~al.} 2010; {Leja}, {van Dokkum}, \& {Franx} 2013). Taken at face value, this is a paradoxical result: where is all of the merging mass going?

 One potential solution lies in the extended light profiles of massive galaxies. Massive galaxies have large, low surface brightness components which in dense environments will blend naturally into the inter-cluster light (ICL); indeed, the most extended objects have luminosities and radii which are comparable to entire galaxy clusters ({Kluge} {et~al.} 2020). As a consequence, standard photometric techniques substantially underestimate both the luminosity and size of massive galaxies, typically by over-estimating the sky subtraction ({Bernardi} {et~al.} 2010). After accounting for these faint, extended components, {Bernardi} {et~al.} (2013) show that the $z\approx0$ integrated stellar mass density increases by 20\% and the number density of galaxies with log(M/M$_{\odot})=11.7$ increases by a factor of 5.

It is unclear whether the extended low surface brightness features around massive galaxies are well-measured in standard photometric catalogs such as those fit here. Specifically, standard photometric apertures are likely too small to encompass all of the light from massive galaxies. For example, {Mowla} {et~al.} (2019) find that galaxies with log(M/M$_{\odot}) >11.3$ have a median effective radius of $\sim9.3$ kpc at $z\sim0.2$. The {Laigle} {et~al.} (2016) catalog uses 3$''$ photometric apertures, corresponding to $\sim$10 kpc at $z=0.2$. This suggests that the standard technique does not directly measure almost half of the light in massive galaxies at low redshifts; indeed, {Mowla} {et~al.} (2019) show in their Appendix that aperture fluxes systematically underestimate fluxes from profile fitting by up to a magnitude for the largest objects. This remains true even when using larger apertures in ground-based photometry; {van Dokkum} {et~al.} (2010) finds that S\'{e}rsic fits to the light profiles of massive galaxies suggest that standard aperture miss 5\% of the flux at z=2, increasing to 15\% at z=0.6. Unfortunately, this offset is also likely to have a redshift dependence, as a fixed angular aperture will capture a smaller fraction of the total galaxy as the redshift decreases.

Thus we caution that the lack of evolution in the massive end observed in this work should not be over-interpreted, until a more complete accounting of the extended light of massive galaxies is performed, especially at $z<0.5$ where the angular size of massive galaxies is comparable to or larger than standard photometric apertures.

\section{Conclusion}
\label{sec:conclusion}
In this paper we present new stellar mass functions over the redshift interval $0.2 < z < 3$. We use the \prospector{} SED-fitting code to infer stellar masses. The inputs are rest-frame UV-IR photometry and measured redshifts from the publicly-available COSMOS-2015 and 3D-HST galaxy catalogs. As shown in {Leja} {et~al.} (2019b), \prospector{} infers $0.1-0.3$ dex larger stellar masses than standard approaches, largely due to the use of nonparametric star formation histories.

We couple these mass measurements with a new methodology for measuring the evolution of the stellar mass function. The standard maximum likelihood approach slices a survey into multiple distinct volumes and fits independent mass functions in each volume. Our new continuity modeling approach constrains the redshift evolution of the mass function using all of the observed masses and redshifts at once, assuming that the mass function evolves smoothly with redshift. It is conditioned on the full stellar mass posteriors (requiring no assumption about the shape of the uncertainties), and includes the effects of sampling variance. We demonstrate that the redshift evolution inferred with this method is more consistent than standard methodology, particularly below the mass-complete limit and at the massive end of the mass function.

The stellar mass function in this work shows higher number densities at a fixed stellar mass than almost any other measurement in the literature, with integrated stellar mass densities $\sim$50\% higher than other studies. This is largely due to differences in SED-fitting methodology: the flexible nonparametric star formation histories used in \prospector{} produce older ages and therefore more massive galaxies than standard approaches. The rate of change of the integrated stellar mass density peaks at $z=1.5$, lower than the consensus model of $z\approx1.9$ ({Madau} \& {Dickinson} 2014). Key areas for future work on the galaxy stellar mass function include folding redshift uncertainties into the model constraints, performing fits to fainter objects in order to constrain the evolution of the low-mass slope, and explaining the apparent lack of evolution in the number density of massive galaxies between $0 < z < 1$.

This paper is the first in a series of three papers which aim to present a unified picture of galaxy assembly between $0.2 < z < 3$ as inferred by \prospector{}. Subsequent papers in this series will address the redshift evolution of the galaxy star-forming main sequence and the overall star formation rate density.

\acknowledgements J.L. is supported by an NSF Astronomy and Astrophysics Postdoctoral Fellowship under award AST-1701487. J.S.S. is partially supported through funding from the Harvard Data Science Initiative. The computations in this paper were run on the Odyssey cluster supported by the FAS Division of Science, Research Computing Group at Harvard University. This research made use of \texttt{astropy}\footnote{http://www.astropy.org}, a community-developed core Python package for Astronomy ({Astropy Collaboration} {et~al.} 2013, 2018). This work is based on data products from observations made with ESO Telescopes at the La Silla Paranal Observatory under ESO program ID 179.A-2005 and on data products produced by TERAPIX and the Cambridge Astronomy Survey Unit on behalf of the UltraVISTA consortium.

\software{\texttt{Prospector} (Johnson \& Leja 2017), \texttt{python-fsps} (Foreman-Mackey, Sick, \&  Johnson 2014), \texttt{Astropy} ({Astropy Collaboration} {et~al.} 2013, 2018), \texttt{FSPS} ({Conroy} {et~al.} 2009), \texttt{matplotlib} (Caswell {et~al.} 2018), \texttt{scipy} ( ), \texttt{ipython} (P\'erez \& Granger 2007), \texttt{numpy} (Walt, Colbert, \& Varoquaux 2011), \texttt{dynesty} ({Speagle} 2020), \texttt{emcee} ({Foreman-Mackey} {et~al.} 2013)}

\appendix
\section{Fitting Mock Data with the Continuity Model}
\subsection{Accurate Treatment of Mass Uncertainty}
\label{sec:mock}
Here we test our methodology by generating mock galaxies with noisy stellar masses, and fitting them with the continuity model.

The galaxies are generated by first assuming an underlying stellar mass function near the recovered posterior values. A mock survey is performed over the angular size of the 3D-HST survey over $0.5 < z < 3$, using the measured 3D-HST mass completeness limits. The combination of the stellar mass function and the areal coverage is used to determine the number of objects, and the (noiseless) redshifts are drawn randomly proportional to the derivative of the total galaxy number density. No additional sampling variance is included in either the mock generation or in the fitting process.

Next, stellar masses are assigned by drawing from the stellar mass function. The observed stellar masses are perturbed from the true mass by drawing from a Student's-$t$ distribution with $\nu=6$ degrees of freedom and a standard deviation of 0.3 dex centered on the true mass. The Student's-$t$ distribution is qualitatively similar to a normal distribution but with wider tails, and is chosen to demonstrate the robustness to non-Gaussian uncertainties. Posterior samples are generated around the perturbed mass using the same Student's-$t$ distribution. These uncertainties are intentionally chosen to be larger than the observed stellar mass uncertainties as a test of the methodology.

\begin{figure*}[t!]
\begin{center}
\includegraphics[width=\linewidth]{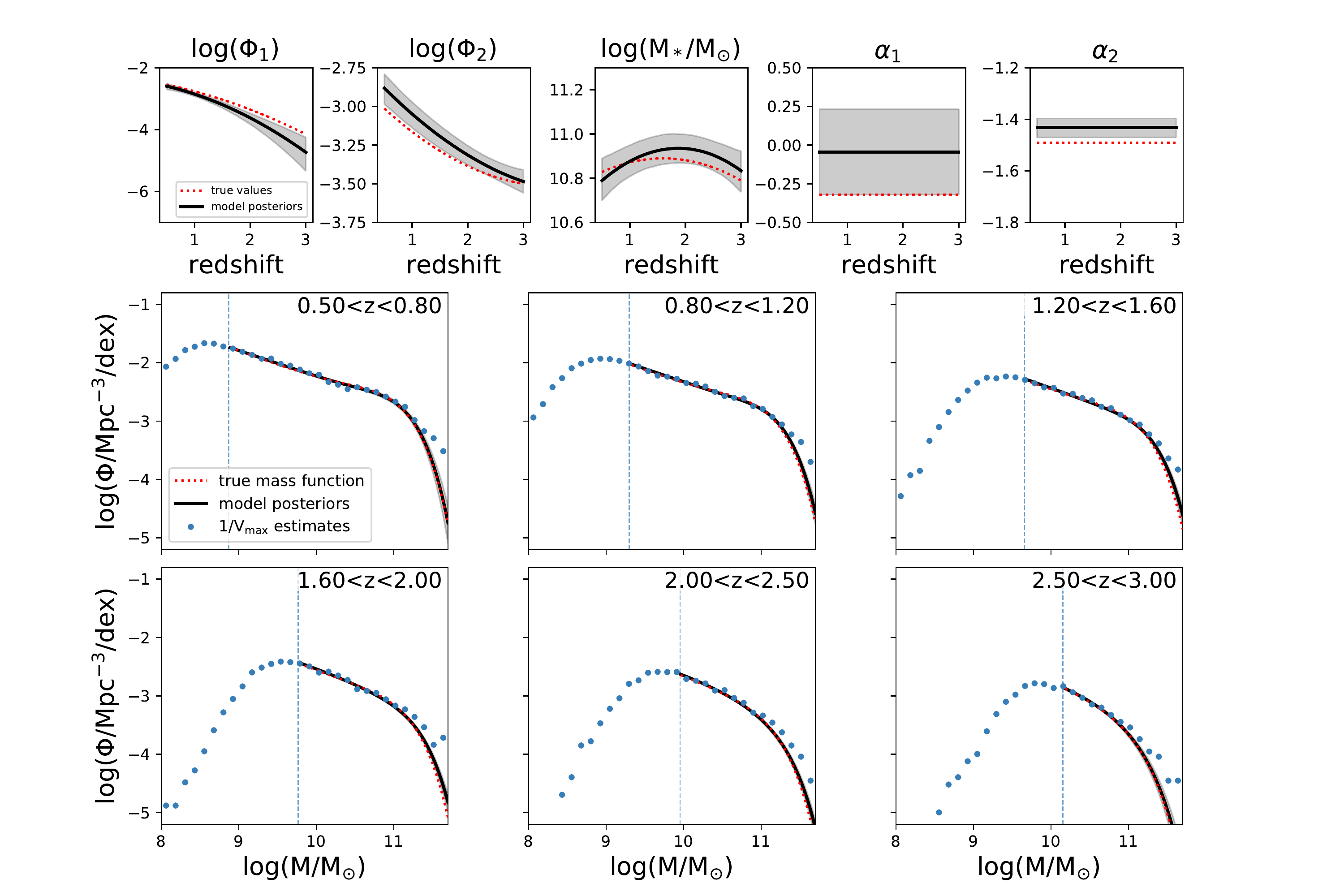}
\caption{Demonstrating the recovery of mock input parameters using the continuity model. The top panels show the redshift evolution of both the input and recovered Schechter parameters, while the bottom panels show the redshift evolution of the mass function in redshift bins. The lines indicate the posterior median values while the shaded regions indicate 2$\sigma$ model uncertainties. The model mass function uncertainties are typically smaller than the line width. The dotted lines indicates the mass-incomplete limit.}
\label{fig:mock}
\end{center}
\end{figure*}

Figure \ref{fig:mock} shows that the continuity model accurately recovers both the input Schechter parameters and the underlying mass function. Notably, it recovers the shape of the massive end even in the face of significant and non-Gaussian stellar mass uncertainties. The 1/V$_{max}$ estimates do not adjust for the effect of stellar mass uncertainties and as a result, overestimate the density of massive galaxies via Eddington bias. The $\sim$1$\sigma$ over-estimate of $\alpha_2$ and $\phi_2$ is because these parameters are strongly covariant (see Figure \ref{fig:corner}). Even though the Schechter parameters do not exactly match the inputs, the posterior predictive number densities agree well with the inputs, which is the primary goal of the fit. Exploring physical models which have fewer degeneracies than a double Schechter is suggested for future work to avoid these issues.

Finally, we caution that this test represents an ideal scenario, where the input mass function evolves smoothly according to the model assumptions and the noise properties are known perfectly. Practical applications of this method are subject to additional unknown systematic effects stemming from differences between the model assumptions and the real universe.

\subsection{Constraining Sampling Variance}
\label{sec:mock_sigma}
Here we explore the ability to constrain the additional sampling variance term in the continuity model. To test this, we fit the continuity model to galaxies from the Horizon-AGN simulation ({Dubois} {et~al.} 2014). As this is a cosmological hydrodynamical simulation, it contains the large-scale structure which the sampling variance term is intended to marginalize over.

{Laigle} {et~al.} (2019) has extracted stellar masses and redshifts from the Horizons-AGN simulation in a light-cone designed to emulate the COSMOS survey. We pass these parameters to the continuity model, assuming no redshift uncertainty and using the same Student's-t distribution of mass uncertainties described in Appendix \ref{sec:mock}, but with a much smaller standard deviation of 0.02 dex. This smaller standard deviation is adopted in order to better isolate the effect of cosmic structure on the posteriors. Galaxies in the range $0.2 < z < 0.8$ are included to emulate the cuts used in the main analysis. The galaxy catalog is complete down to $10^9$ M$_{\odot}$ in {\it total mass formed}; accordingly we adopt a completeness limit of $10^9$ M$_{\odot}$ in stellar mass.

\begin{figure*}[t!]
\begin{center}
\includegraphics[width=\linewidth]{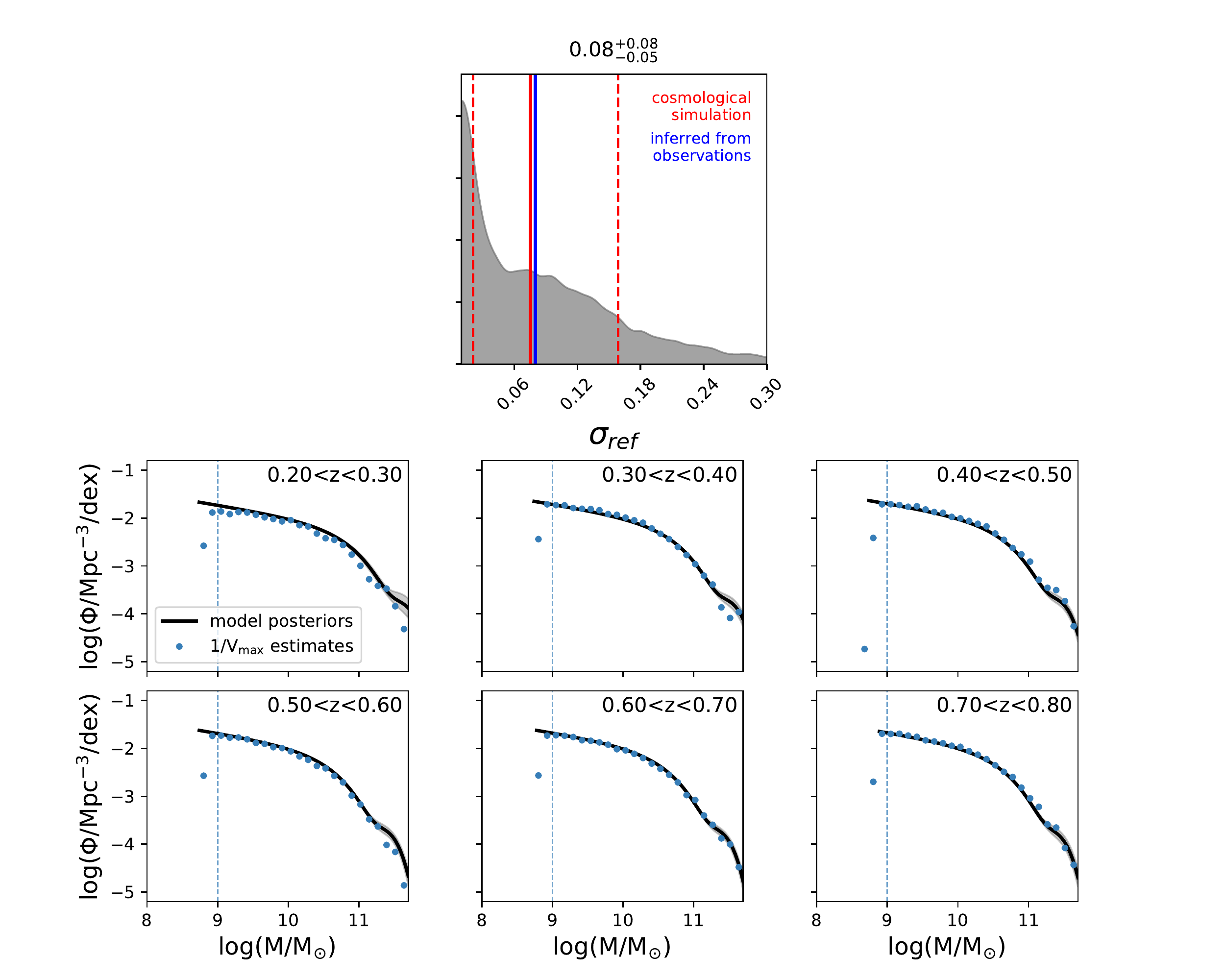}
\caption{Exploring the recovery of the sampling variance term when fitting a cosmological simulation with the continuity model. The top panel shows the marginalized posterior for $\sigma_{\mathrm{ref}}$; the median and 1$\sigma$ contours are marked in red, while the value inferred from the observations is marked in blue.
While the posterior from fitting the simulations is largely prior-dominated, it contains a peak near $\sigma_{\mathrm{ref}}$ = 0.08, similar to the observational result. This suggests the sampling variance term is properly recovering the variance from large-scale structure. The bottom panels show the redshift evolution of the mass function in redshift bins. The lines indicate the posterior median values while the shaded regions indicate 2$\sigma$ model uncertainties. The dotted lines indicates the mass completeness limit.}
\label{fig:mock_cosmos}
\end{center}
\end{figure*}

The resulting fit is shown in Figure \ref{fig:mock_cosmos}. The posterior for $\sigma_{ref}$ retains a clear signature of the adopted logarithmic prior, but with a large bump around $\sigma_{ref} = 0.08$, similar to the value inferred from the observations. While there is not enough information to uniquely identify the magnitude of the additional sampling variance, both the observational and the simulated data prefer a similar value. This suggests the model is marginalizing over the correct magnitude of the sampling variance term. A stronger, more informative prior on the distribution of large-scale structure in the universe would have very little effect on the results of this study, as can be seen by the lack of significant covariance between $\sigma_{\mathrm{ref}}$ and the mass function parameters in Figure \ref{fig:corner}.

The simulated mass function has an excess over a typical Schechter mass function at very high masses, visible in Figure \ref{fig:mock_cosmos}; the upper limit for $\alpha_1$ was increased to 6 in order to accurately describe the mean evolution of the massive end of the simulated mass function.

\section{Guide to generating a mass function using the continuity model parameters}
\label{sec:guide}
Here we demonstrate how to generate a mass function and associated uncertainties at some redshift $z_0$ using the parameters of the continuity model. This code can also be adapted to sample the posterior for other purposes; for example, calculating the integrated stellar mass density using equation \ref{eqn:smd}. The fit parameters and their 1$\sigma$ marginalized uncertainties are available in Figure \ref{fig:corner}. 

The first step is to convert the redshift-dependent parameters into quadratic coefficients. The model parameters are three points on a quadratic line, corresponding to $z_1=3.0$, $z_2=1.6$, and $z_3=0.2$. Using the associated $y_1$, $y_2$, and $y_3$ values one can convert to quadratic coefficients via
\begin{align}
    a &= \frac{(y_3 - y_1) + (y_2-y_1)\frac{z_1-z_3}{z_2-z_1}}
             {z_3^2 - z_1^2 + \frac{(z_2^2-z_1^2)(z_1-z_3)}{z_2-z_1}} \\
    b & = \frac{y_2 - y_1 - a(z_2^2-z_1^2)}{(z_2-z_1)} \\
    c & = y_1 - az_1^2 - bz_1
\end{align}
for a quadratic defined as
\begin{equation}
    y(z)=az^2 + bz + c
\end{equation}
Using this, the redshift-dependent parameters $\phi_1$, $\phi_2$, and $\mathcal{M_*}$ can be calculated at an arbitrary redshift $z_0$, where $z_0$ is bounded such that $0.2 < z_0 < 3$. These parameters can then be inserted into equation (\ref{eqn:dblschech}) to construct the stellar mass function.

Precisely generating the uncertainties in the mass function requires access to the posterior samples. In practice, these can be simulated without much loss of precision by assuming uncorrelated Gaussian uncertainties. Below is a section of \texttt{python} code which generates a posterior median mass function from the continuity model parameters and their associated 1$\sigma$ uncertainties.

\begin{verbatim}
import numpy as np

def schechter(logm, logphi, logmstar, alpha, m_lower=None):
    """
    Generate a Schechter function (in dlogm).
    
    """

    phi = ((10**logphi) * np.log(10) *
           10**((logm - logmstar) * (alpha + 1)) *
           np.exp(-10**(logm - logmstar)))

    return phi

def parameter_at_z0(y,z0,z1=0.2,z2=1.6,z3=3.0):
    """ 
    Compute parameter at redshift `z0` as a function
    of the polynomial parameters `y` and the
    redshift anchor points `z1`, `z2`, and `z3`.
    
    """

    y1, y2, y3 = y
    a = (((y3 - y1) + (y2 - y1) / (z2 - z1) * (z1 - z3)) /
         (z3**2 - z1**2 + (z2**2 - z1**2) / (z2 - z1) * (z1 - z3)))
    b = ((y2 - y1) - a * (z2**2 - z1**2)) / (z2 - z1)
    c = y1 - a * z1**2 - b * z1
    return a * z0**2 + b * z0 + c

# Continuity model median parameters + 1-sigma uncertainties.
pars = {'logphi1': [-2.44, -3.08, -4.14],
        'logphi1_err': [0.02, 0.03, 0.1],
        'logphi2': [-2.89, -3.29, -3.51],
        'logphi2_err': [0.04, 0.03, 0.03],
        'logmstar': [10.79,10.88,10.84],
        'logmstar_err': [0.02, 0.02, 0.04],
        'alpha1': [-0.28],
        'alpha1_err': [0.07],
        'alpha2': [-1.48],
        'alpha2_err': [0.1]}

# Draw samples from posterior assuming independent Gaussian uncertainties.
# Then convert to mass function at `z=z0`.
draws = {}
ndraw = 1000
z0 = 1.0
for par in ['logphi1', 'logphi2', 'logmstar', 'alpha1', 'alpha2']:
    samp = np.array([np.random.normal(median,scale=err,size=ndraw)
                     for median, err in zip(pars[par], pars[par+'_err'])])
    if par in ['logphi1', 'logphi2', 'logmstar']:
      draws[par] = parameter_at_z0(samp,z0)
    else:
      draws[par] = samp.squeeze()

# Generate Schechter functions.
logm = np.linspace(8, 12, 100)[:, None]  # log(M) grid
phi1 = schechter(logm, draws['logphi1'],  # primary component
                 draws['logmstar'], draws['alpha1'])
phi2 = schechter(logm, draws['logphi2'],  # secondary component
                 draws['logmstar'], draws['alpha2'])
phi = phi1 + phi2  # combined mass function

# Compute median and 1-sigma uncertainties as a function of mass.
phi_50, phi_84, phi_16 = np.percentile(phi, [50, 84, 16], axis=1)
\end{verbatim}


\end{document}